\documentclass[12pt,fleqn]{book} 

\usepackage[top=3cm,bottom=3cm,left=3.2cm,right=3.2cm,headsep=10pt,letterpaper]{geometry} 

\usepackage{xcolor} 
\definecolor{ocre}{RGB}{100,70,201} 

\usepackage{avant} 
\usepackage{mathptmx} 

\usepackage{siunitx}

\usepackage{microtype} 
\usepackage[utf8]{inputenc} 
\usepackage[T1]{fontenc} 
\usepackage{rotating}

\usepackage{hyperref}
\usepackage{footnote}

\usepackage{hyperref}
\hypersetup{
    colorlinks=true,
    linkcolor=blue,
    filecolor=magenta,      
    urlcolor=cyan,
    pdftitle={Overleaf Example},
    pdfpagemode=FullScreen,
    }

\usepackage[nonumberlist,sort=standard]{glossaries}

\usepackage{multirow}
\makeglossaries

\usepackage{titlesec} 

\usepackage{graphicx} 
\graphicspath{{Pictures/}} 

\usepackage{lipsum} 

\usepackage{tikz} 

\usepackage[english]{babel} 

\usepackage{enumitem} 
\setlist{nolistsep} 

\usepackage{booktabs} 

\usepackage{eso-pic} 

\usepackage{titletoc} 

\contentsmargin{0cm} 
\titlecontents{chapter}[1.25cm] 
{\addvspace{15pt}\large\sffamily\bfseries} 
{\color{ocre!60}\contentslabel[\Large\thecontentslabel]{1.25cm}\color{ocre}} 
{}  
{\color{ocre!60}\normalsize\sffamily\bfseries\;\titlerule*[.5pc]{.}\;\thecontentspage} 
\titlecontents{section}[1.25cm] 
{\addvspace{5pt}\sffamily\bfseries} 
{\contentslabel[\thecontentslabel]{1.25cm}} 
{}
{\sffamily\hfill\color{black}\thecontentspage} 
[]
\titlecontents{subsection}[1.25cm] 
{\addvspace{1pt}\sffamily\small} 
{\contentslabel[\thecontentslabel]{1.25cm}} 
{}
{\sffamily\;\titlerule*[.5pc]{.}\;\thecontentspage} 
[] 

\titlecontents{lsection}[0em] 
{\footnotesize\sffamily} 
{}
{}
{}

\titlecontents{lsubsection}[.5em] 
{\normalfont\footnotesize\sffamily} 
{}
{}
{}
 
\usepackage{fancyhdr} 

\fancypagestyle{plain}{\fancyhead{}} 

\makeatletter
\renewcommand{\cleardoublepage}{
\clearpage\ifodd\c@page\else
\hbox{}
\vspace*{\fill}
\thispagestyle{empty}
\newpage
\fi}

\usepackage{amsmath,amsfonts,amssymb,amsthm} 

\newtheoremstyle{ocrenumbox}
{0pt}
{0pt}
{\normalfont}
{}
{\small\bf\sffamily\color{ocre}}
{\;}
{0.25em}
{\small\sffamily\color{ocre}\thmname{#1}\nobreakspace\thmnumber{\@ifnotempty{#1}{}\@upn{#2}}
\thmnote{\nobreakspace\the\thm@notefont\sffamily\bfseries\color{black}---\nobreakspace#3.}} 

\newtheoremstyle{blacknumex}
{5pt}
{5pt}
{\normalfont}
{} 
{\small\bf\sffamily}
{\;}
{0.25em}
{\small\sffamily{\tiny\ensuremath{\blacksquare}}\nobreakspace\thmname{#1}\nobreakspace\thmnumber{\@ifnotempty{#1}{}\@upn{#2}}
\thmnote{\nobreakspace\the\thm@notefont\sffamily\bfseries---\nobreakspace#3.}}

\newtheoremstyle{blacknumbox} 
{0pt}
{0pt}
{\normalfont}
{}
{\small\bf\sffamily}
{\;}
{0.25em}
{\small\sffamily\thmname{#1}\nobreakspace\thmnumber{\@ifnotempty{#1}{}\@upn{#2}}
\thmnote{\nobreakspace\the\thm@notefont\sffamily\bfseries---\nobreakspace#3.}}

\newtheoremstyle{ocrenum}
{5pt}
{5pt}
{\normalfont}
{}
{\small\bf\sffamily\color{ocre}}
{\;}
{0.25em}
{\small\sffamily\color{ocre}\thmname{#1}\nobreakspace\thmnumber{\@ifnotempty{#1}{}\@upn{#2}}
\thmnote{\nobreakspace\the\thm@notefont\sffamily\bfseries\color{black}---\nobreakspace#3.}} 
\makeatother

\newcounter{dummy} 
\numberwithin{dummy}{section}
\theoremstyle{ocrenumbox}
\newtheorem{theoremeT}[dummy]{Theorem}

\newtheorem{exerciseT}{Exercise}[chapter]
\theoremstyle{blacknumex}
\newtheorem{exampleT}{Example}[chapter]
\theoremstyle{blacknumbox}

\newtheorem{definitionT}{Definition}[section]
\newtheorem{corollaryT}[dummy]{Corollary}
\theoremstyle{ocrenum}

\RequirePackage[framemethod=default]{mdframed} 

\newmdenv[skipabove=7pt,
skipbelow=7pt,
backgroundcolor=black!5,
linecolor=ocre,
innerleftmargin=5pt,
innerrightmargin=5pt,
innertopmargin=5pt,
leftmargin=0cm,
rightmargin=0cm,
innerbottommargin=5pt]{tBox}

\newmdenv[skipabove=7pt,
skipbelow=7pt,
rightline=false,
leftline=true,
topline=false,
bottomline=false,
backgroundcolor=ocre!10,
linecolor=ocre,
innerleftmargin=5pt,
innerrightmargin=5pt,
innertopmargin=5pt,
innerbottommargin=5pt,
leftmargin=0cm,
rightmargin=0cm,
linewidth=4pt]{eBox}	

\newmdenv[skipabove=7pt,
skipbelow=7pt,
rightline=false,
leftline=true,
topline=false,
bottomline=false,
linecolor=ocre,
innerleftmargin=5pt,
innerrightmargin=5pt,
innertopmargin=0pt,
leftmargin=0cm,
rightmargin=0cm,
linewidth=4pt,
innerbottommargin=0pt]{dBox}	

\newmdenv[skipabove=7pt,
skipbelow=7pt,
rightline=false,
leftline=true,
topline=false,
bottomline=false,
linecolor=gray,
backgroundcolor=black!5,
innerleftmargin=5pt,
innerrightmargin=5pt,
innertopmargin=5pt,
leftmargin=0cm,
rightmargin=0cm,
linewidth=4pt,
innerbottommargin=5pt]{cBox}

\makeatletter
\renewcommand{\@seccntformat}[1]{\llap{\textcolor{ocre}{\csname the#1\endcsname}\hspace{1em}}}                    
\renewcommand{\section}{\@startsection{section}{1}{\z@}
{-4ex \@plus -1ex \@minus -.4ex}
{1ex \@plus.2ex }
{\normalfont\large\sffamily\bfseries}}
\renewcommand{\subsection}{\@startsection {subsection}{2}{\z@}
{-3ex \@plus -0.1ex \@minus -.4ex}
{0.5ex \@plus.2ex }
{\normalfont\sffamily\bfseries}}
\renewcommand{\subsubsection}{\@startsection {subsubsection}{3}{\z@}
{-2ex \@plus -0.1ex \@minus -.2ex}
{.2ex \@plus.2ex }
{\normalfont\small\sffamily\bfseries}}                        
\renewcommand\paragraph{\@startsection{paragraph}{4}{\z@}
{-2ex \@plus-.2ex \@minus .2ex}
{.1ex}
{\normalfont\small\sffamily\bfseries}}

\usepackage{hyperref}
\hypersetup{hidelinks,backref=true,pagebackref=true,hyperindex=true,colorlinks=false,breaklinks=true,urlcolor= ocre,bookmarks=true,bookmarksopen=false,pdftitle={Title},pdfauthor={Author}}

\newcommand{\thechapterimage}{}
\newcommand{\chapterimage}[1]{\renewcommand{\thechapterimage}{#1}}

\def\thechapter{\arabic{chapter}}
\def\@makechapterhead#1{
\thispagestyle{empty}
{\centering \normalfont\sffamily
\ifnum \c@secnumdepth >\m@ne
\if@mainmatter
\startcontents
\begin{tikzpicture}[remember picture,overlay]
\node at (current page.north west)
{\begin{tikzpicture}[remember picture,overlay]
\node[anchor=north west,inner sep=0pt] at (0,0) {\includegraphics[width=\paperwidth]{\thechapterimage}};
\draw[anchor=west] (5cm,-9cm) node [rounded corners=20pt,fill=ocre!10!white,text opacity=1,draw=ocre,draw opacity=1,line width=1.5pt,fill opacity=.6,inner sep=12pt]{\huge\sffamily\bfseries\textcolor{black}{\thechapter. #1\strut\makebox[22cm]{}}};
\end{tikzpicture}};
\end{tikzpicture}}
\par\vspace*{230\p@}
\fi
\fi}

\def\@makeschapterhead#1{
\thispagestyle{empty}
{\centering \normalfont\sffamily
\ifnum \c@secnumdepth >\m@ne
\if@mainmatter
\begin{tikzpicture}[remember picture,overlay]
\node at (current page.north west)
{\begin{tikzpicture}[remember picture,overlay]
\node[anchor=north west,inner sep=0pt] at (0,0) {\includegraphics[width=\paperwidth]{\thechapterimage}};
\draw[anchor=west] (5cm,-9cm) node [rounded corners=20pt,fill=ocre!10!white,fill opacity=.6,inner sep=12pt,text opacity=1,draw=ocre,draw opacity=1,line width=1.5pt]{\huge\sffamily\bfseries\textcolor{black}{#1\strut\makebox[22cm]{}}};
\end{tikzpicture}};
\end{tikzpicture}}
\par\vspace*{230\p@}
\fi
\fi
}
\makeatother 

\begin{document}

\pagestyle{fancy}
\begingroup
\thispagestyle{empty}
\AddToShipoutPicture*{\put(0,0){\includegraphics[scale=1]
{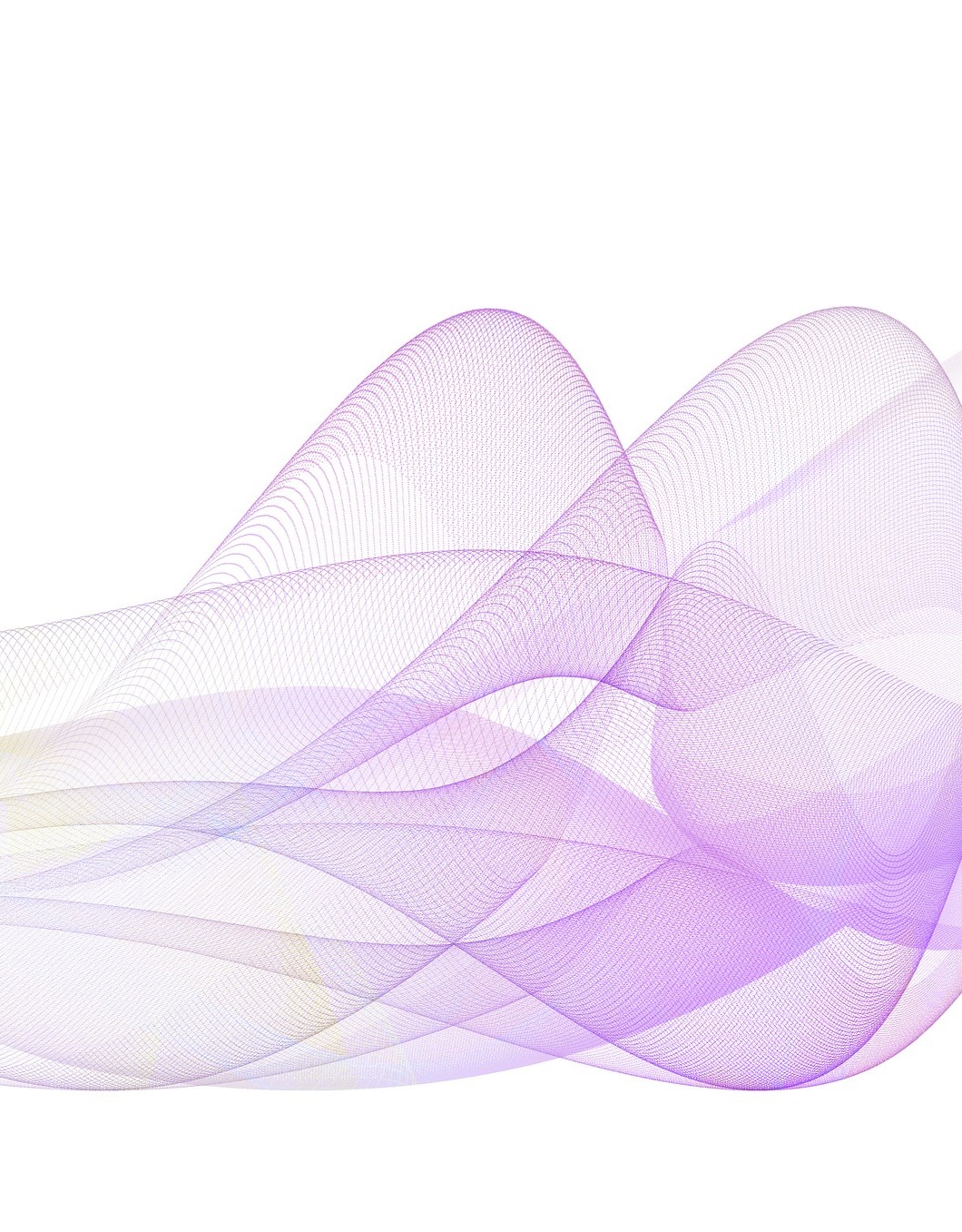}}} 
\centering
\vspace*{5cm}
\par\normalfont\fontsize{35}{35}\sffamily\selectfont
\textbf{Report on the Advanced Linear Collider Study Group (ALEGRO) Workshop 2024}\\
\vspace*{1cm}
{\huge ALEGRO 2024}\par 
\endgroup


\newpage
~\vfill
\thispagestyle{empty}

\noindent Copyright \copyright\ 2024 \\ 

\noindent \textsc{Report on the Advanced Linear Collider Study Group Workshop}\\

\noindent \href{URL https://indico.cern.ch/event/1364999/}{\textsc{URL https://indico.cern.ch/event/1364999/}
}\\ 


\noindent This report is the outcome of the ALEGRO2024 workshop held in March 2024 at IST Lisbon  and supported by IST.\\ 

\noindent \textit{First release, date: \today} 


\newpage

\begingroup
\thispagestyle{empty}
\centering
\vspace*{5cm}
\par\normalfont\fontsize{35}{35}\sffamily\selectfont
\textbf{Report on the Advanced Linear Collider Study Group (ALEGRO) Workshop 2024}\\
\vspace*{0.5cm}
{\Huge Jorge Vieira}\par 
\vspace{0.3cm}
\large{Chair of the ALEGRO 2024 workshop}\\
\vspace{0.5cm}
{\Huge Brigitte Cros, Patric Muggli}\par 
\vspace{0.3cm}
\large{for the ICFA Advanced and Novel Accelerators Panel}\\
\endgroup
\newpage

\thispagestyle{empty}
\vspace*{5cm}










{\large Let us make our future now, and let us make our dreams tomorrow's reality.
\vspace{0.5cm}

Malala Yousafzai, Peace Nobel Prize Winner, 2014}



\newpage
\thispagestyle{empty}
 \begin{figure}[h]
	\begin{center}	
\includegraphics[width=7cm]{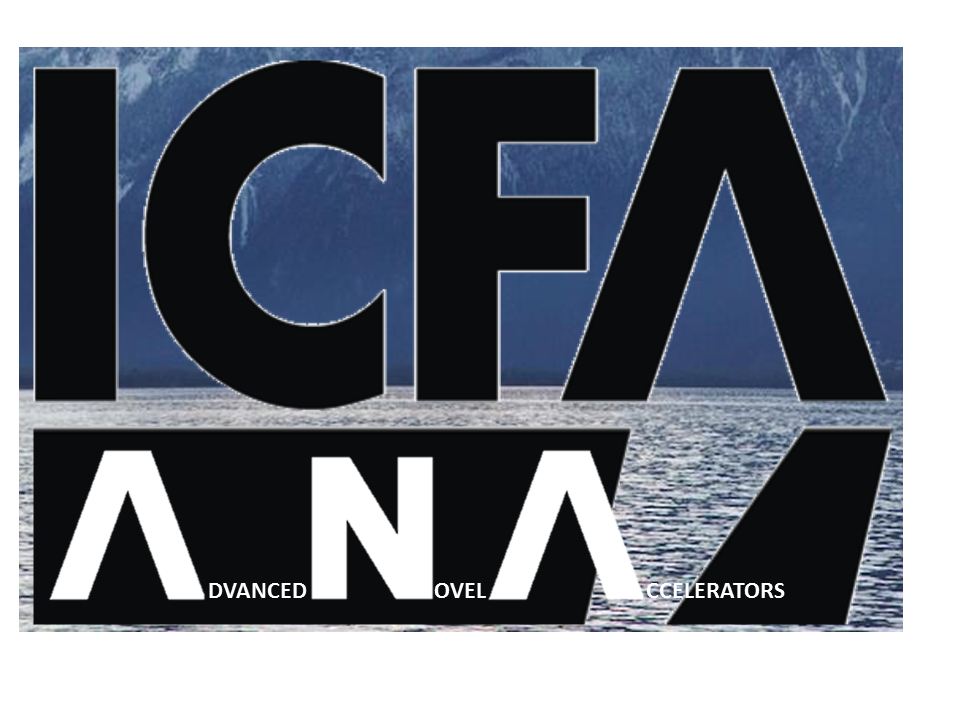}				\end{center}
\end{figure}
\begingroup
\par\normalfont\fontsize{20}{20}\sffamily\selectfont
\hspace*{2cm} \textbf{Executive Summary}\\
\endgroup
The workshop focused on the application of ANAs to %
particle physics
\newglossaryentry{PP}{name={PP},description={Particle Physics}}  (PP) and %
high-energy physics
\newglossaryentry{HEP}{name={HEP},description={High-Energy Physics}}  (HEP), %
keeping in mind the ultimate goal of a collider at the energy frontier (10\,TeV, e$^+$/e$^-$, e$^-$/e$^-$, or $\gamma\gamma$). %
\newglossaryentry{e-}{name={e$^-$},description={electron}}
\newglossaryentry{e+}{name={e$^+$},description={positron}}
\newglossaryentry{p+}{name={p$^+$},description={proton}}
\newglossaryentry{IFEL}{name={IFEL},description={Inverse Free Electron Laser}}
\newglossaryentry{ICFA}{name={ICFA},description={International Committee for Future Accelerators}}
\newglossaryentry{ANA}{name={ANA},description={Advanced and Novel Accelerator}}

The development of ANAs is conducted at universities and national laboratories worldwide. %
The community is thematically broad and diverse, in particular since lasers suitable for ANA research (multi-hundred-terawatt peak power, a few tens of femtosecond-long pulses) and acceleration of electrons to hundreds of mega electron volts to multi giga electron volts became commercially available. %
The community spans  several continents (Europe, America, Asia), including more than 62 laboratories in more than 20 countries. %

It is among the missions of the ICFA-ANA panel to feature the amazing progress made with ANAs, to provide international coordination and to foster international collaborations towards a future HEP collider. %

The scope of this edition of the workshop was to discuss the recent progress and necessary steps towards realizing a linear collider for particle physics based on novel-accelerator technologies (laser or beam driven in plasma or structures). %
Updates on the relevant aspects of the European Strategy for Particle Physics (ESPP) Roadmap Process as well as of the P5 (in the US) were presented, and ample time was dedicated to discussions. %
\newglossaryentry{ESPP}{name={ESPP},description={European Strategy for Particle Physics}}
\newglossaryentry{P5}{name={P5},description={Particle Physics Project Prioritization Panel}}
The major outcome of the workshop is the decision for ALEGRO to coordinate efforts in Europe, in the US, and in Asia towards a pre-CDR for an ANA-based, 10\,TeV CM collider. %
This goal of this coordination is to lead to a funding proposal to be submitted to both EU and EU/US funding agencies.

This document presents a summary of the workshop, as seen by the co-chairs, as well as short 'one-pagers' written by the presenters at the workshop. %

\vspace{1cm}
Brigitte Cros, Patric Muggli, Jorge Vieira

On behalf of the ICFA-ANA panel

\newpage
\vspace{0.5cm}
\noindent \textbf{Members of the ICFA ANA panel, March 2024}
\\
\\
\noindent Bruce Carlsten, \textit{Los Alamos National Laboratory (LANL), USA}

\noindent Brigitte Cros, \textit{Centre National de la Recherche Scientifique (CNRS) -- Universit\'{e} Paris Sud, France}

\noindent Massimo Ferrario, \textit{Istituto Nazionale di Fisica Nucleare (INFN), Italy}

\noindent Simon Hooker, \textit{ University of Oxford, UK}

\noindent Tomonao Hosokai, \textit{Univ. Osaka, Japan}

\noindent Masaki Masaki Kando, \textit{National Institutes for Quantum and Radiological Science and Technology, Japan}

\noindent Patric Muggli, \textit{Max Planck Institute for Physics (MPP), Germany}

\noindent Jens Osterhoff, \textit{Lawrence Berkeley National Laboratory (LBNL), USA}

\noindent Philippe Piot, \textit{Northern Illinois University (NIU), Fermi National Accelerator Laboratory (FNAL), USA}

\noindent James Rosenzweig, \textit{University of California, Los Angeles (UCLA), USA}

\noindent Carl Schroeder, \textit{Lawrence Berkeley National Laboratory (LBNL),  USA}

\noindent Chuanxiang Tang (chair), \textit{Tsinghua University, China}


\newpage
\newglossaryentry{CERN}{name={CERN},description={European Organization for Nuclear Research}}
\newglossaryentry{SLAC}{name={SLAC},description={Stanford Linear Accelerator Center}}
\newglossaryentry{LBNL}{name={LBNL},description={Lawrence Berkeley National Laboratory}}
\newglossaryentry{DESY}{name={DESY},description={Deutsches Elektronen-SYnchrotron}}
\newglossaryentry{INFN}{name={INFN},description={Istituto Nazionale di Fisica Nucleare}}
\newglossaryentry{RD}{name={R\&D},description={Research and Development}}
\newglossaryentry{CLIC}{name={CLIC},description={Compact LInear Collider}}

\newpage
\vspace{0.5cm}
\noindent \textbf{List of co-authors in alphabetical order}
\\
\\
I.A.~Andriyash, \textit{Laboratoire d’Optique Appliquée, ENSTA Paris, CNRS, Ecole Polytechnique, Institut Polytechnique de Paris, 828 Bd des Maréchaux, 91762 Palaiseau, France}\\
O.~Apsimon, \textit{University of Manchester, Manchester, U.K.}\\
M.~Backhouse, \textit{The John Adams Institute for Accelerator Science, Imperial College, London SW7 2BZ, United Kingdom} \\
C.~Benedetti, \textit{BELLA Center, Lawrence Berkeley National Laboratory, Berkeley, CA, USA}\\
S.~S.~Bulanov, \textit{BELLA Center, Lawrence Berkeley National Laboratory, Berkeley, CA, USA}\\
A.~Caldwell, \textit{Max-Planck-Institut für Physik, München, Germany}\\
Min Chen, \textit{Key Laboratory for Laser Plasmas, School of Physics and Astronomy, Shanghai Jiao Tong University, Shanghai 200240, China}\\
V.~Cilento, \textit{CERN, Geneva, Switzerland}\\
S.~Corde, \textit{LOA, ENSTA Paris, CNRS, Ecole Polytechnique, Institut Polytechnique de Paris, 91762, Palaiseau, France}\\
R.~D'Arcy, \textit{Oxford University, Oxford, United Kingdom}\\
S.~Diederichs, \textit{Deutsches Elektronen-Synchrotron DESY, 22607 Hamburg, Germany}\\
E.~Ericson, \textit{Paul Scherrer Institute, École Polytechnique Fédérale de Lausanne, Wurenlingen, Switzerland}\\
E.~Esarey, \textit{BELLA Center, Lawrence Berkeley National Laboratory, Berkeley, CA, USA}\\
J.~Farmer, \textit{Max-Planck-Institut für Physik, München, Germany}\\
L.~Fedeli, \textit{Université Paris-Saclay, CEA, LIDYL, 91191 Gif-sur-Yvette, France}\\
A.~Formenti, \textit{Lawrence Berkeley National Laboratory, Berkeley, USA}\\
B.~Foster, \textit{DESY, Hamburg, Germany}\\
M.~Garten, \textit{Lawrence Berkeley National Laboratory, Berkeley, USA}\\
C.G.R.~Geddes, \textit{BELLA Center, Lawrence Berkeley National Laboratory, Berkeley, CA, USA}\\
T.~Grismayer, \textit{GoLP/Instituto de Plasmas e Fus\~{a}o Nuclear, Instituto Superior T\'{e}cnico, Universidade de Lisboa, 1049-001 Lisboa, Portugal}\\
M.~J.~Hogan, \textit{SLAC National Accelerator Laboratory, Menlo Park, CA, USA}\\
S.~Hooker, \textit{University of Oxford, United Kingdom}\\
A.~Huebl, \textit{Lawrence Berkeley National Laboratory LBNL, Berkeley, California 94720, USA}\\
S.~Jalas, \textit{Deutsches Elektronen-Synchrotron DESY, Notkestr.~85, 22607 Hamburg, Germany}\\
M.~Kirchen, \textit{Deutsches Elektronen-Synchrotron DESY, Notkestr.~85, 22607 Hamburg, Germany}\\
R.~Lehe, \textit{Lawrence Berkeley National Laboratory LBNL, Berkeley, California 94720, USA}\\
W.~Leemans, \textit{Accelerator Division, DESY, Hamburg, Germany}\\
Boyuan Li, \textit{Key Laboratory for Laser Plasmas, School of Physics and Astronomy, Shanghai Jiao Tong University, Shanghai 200240, China}\\
C.~A.~Lindstr{\"o}m, \textit{University of Oslo, Oslo, Norway}\\
R.~Losito, \textit{CERN, Geneva, Switzerland}\\
C.~E.~Mitchell, \textit{Lawrence Berkeley National Laboratory, Berkeley, USA}\\
W.~B.~Mori, \textit{University of California, Los Angeles CA, USA}\\
P.~Muggli, \textit{Max Planck Institute for Physics, 80805 Munich, Germany}\\
Z.~Najmudin, \textit{The John Adams Institute for Accelerator Science, Imperial College, London SW7 2BZ, United Kingdom}\\
J.~Osterhoff, \textit{BELLA Center, Lawrence Berkeley National Laboratory, Berkeley, CA, USA \& Deutsches Elektronen-Synchrotron DESY, 22607 Hamburg, Germany}\\
R.~Pattathil, \textit{Central Laser Facility, STFC Rutherford Appleton Laboratory, UKRI, United Kingdom}\\
P.~Piot, \textit{Northern Illinois University, DeKalb Illinois 60115, USA \& Argonne National Laboratory, Argonne, Illinois 60439, USA}\\
K.~P\~oder, \textit{Deutsches Elektronen-Synchrotron DESY, Notkestr.~85, 22607 Hamburg, Germany}\\
A.~F.~Pousa, \textit{Deutsches Elektronen-Synchrotron DESY, 22607 Hamburg, Germany}\\
J.~G.~Power, \textit{Argonne National Laboratory, Argonne, Illinois 60439, USA}\\
A.~Pukhov, \textit{Heinrich-Heine-Universität Düsseldorf, D{\"u}sseldorf, Germany}\\
R.~T.~Sandberg, \textit{Lawrence Berkeley National Laboratory, Berkeley, USA}\\
G.~Sarri, \textit{Queen's University, Belfast, UK}\\
R.~J.~Shalloo, \textit{Deutsches Elektronen-Synchrotron DESY, Notkestr.~85, 22607 Hamburg, Germany}\\
O.~Shapoval, \textit{Lawrence Berkeley National Laboratory, Berkeley, USA}\\
S.~Sch{\"o}bel, \textit{Helmholtz-Zentrum Dresden – Rossendorf, Germany}\\
L.~O.~Silva, \textit{GoLP/Instituto de Plasmas e Fus\~{a}o Nuclear, Instituto Superior T\'{e}cnico, Universidade de Lisboa, 1049-001 Lisboa, Portugal}\\
A.~Sinn, \textit{Deutsches Elektronen-Synchrotron DESY, 22607 Hamburg, Germany \& Universit{\"a}t Hamburg UHH, Mittelweg 177, 20148 Hamburg, Germany}\\
C.~B.~Schroder, \textit{BELLA Center, Lawrence Berkeley National Laboratory, Berkeley, CA, USA \& Department of Nuclear Engineering, University of California, Berkeley, California 94720, USA}\\
Zhengming Sheng, \textit{Key Laboratory for Laser Plasmas, School of Physics and Astronomy, Shanghai Jiao Tong University, Shanghai 200240, China \& Tsung-Dao Lee Institute, Shanghai Jiao Tong University, Shanghai 200240, China}\\
D.~Terzani, \textit{BELLA Center, Lawrence Berkeley National Laboratory, Berkeley, CA, USA}\\
M.~Thévenet, \textit{Deutsches Elektronen-Synchrotron DESY, 22607 Hamburg, Germany}\\
M.~Turner, \textit{CERN, Geneva, Switzerland}\\
J.-L.~Vay, \textit{Lawrence Berkeley National Laboratory LBNL, Berkeley, California 94720, USA}\\
J.~Vieira, \textit{GoLP/Instituto de Plasmas e Fus\~{a}o Nuclear, Instituto Superior T\'{e}cnico, Universidade de Lisboa, 1049-001 Lisboa, Portugal}\\
D.~Völker, \textit{DESY, Hamburg, Germany}\\
Jie Zhang, \textit{Key Laboratory for Laser Plasmas, School of Physics and Astronomy, Shanghai Jiao Tong University, Shanghai 200240, China \& Tsung-Dao Lee Institute, Shanghai Jiao Tong University, Shanghai 200240, China}\\
W.~Zhang, \textit{GoLP/Instituto de Plasmas e Fus\~{a}o Nuclear, Instituto Superior T\'{e}cnico, Universidade de Lisboa, 1049-001 Lisboa, Portugal}\\
Xinzhe Zhu, \textit{Key Laboratory for Laser Plasmas, School of Physics and Astronomy, Shanghai Jiao Tong University, Shanghai 200240, China}\\

\vspace{2.0cm}
\noindent \textbf{Acknowledgments}
\\
\\
\noindent This report is the result of the work of many people, including the ICFA-ANA panel members, the workshop organizing committee, and the workshop participants. %
In addition, the contribution of the IST administrative staff (Claudia Rom\~ao and Susana Muinos) and Local Organizing Committee (Jorge Vieira, Bernardo Malaca, Camilla Willim, and Rafael Almeida) is also acknowledged and greatly appreciated. \\%
\newpage

\chapterimage{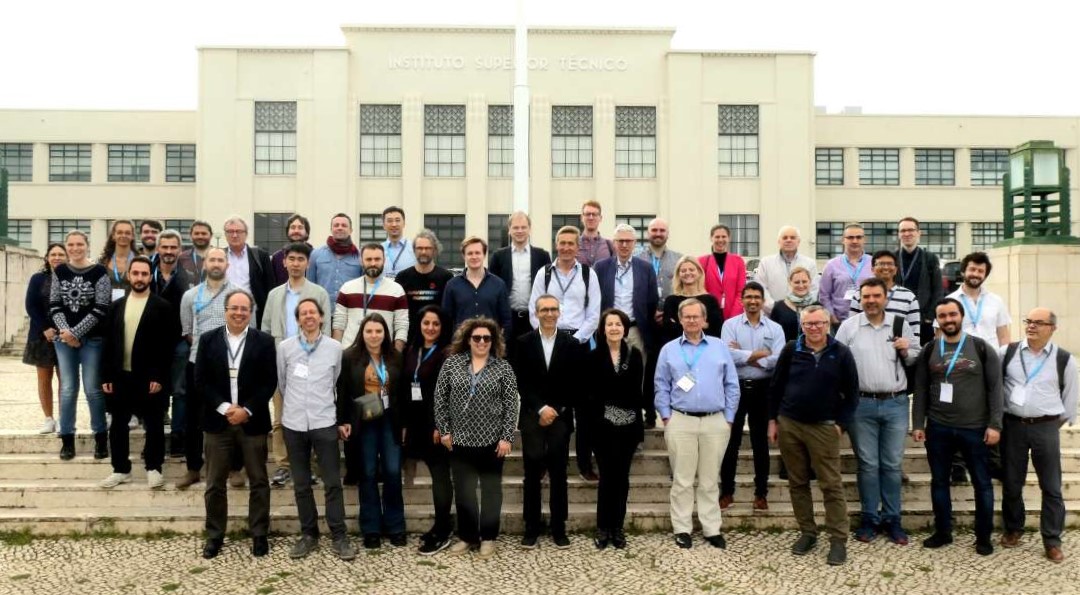} 

\pagestyle{empty} 

\tableofcontents 

\cleardoublepage 


\pagestyle{fancy} 


\chapterimage{images_head_plantp}
\chapter{Overview of the workshop}


We provide here our \textit{brief overview} of the workshop. %
We refer to the name of the presenters, though the views expressed are ours. %
Each speaker was invited to submit a "one-pager" about her/his presentation. %
These are available in the next chapter, in the order of presentation at the workshop. %

\section{Purpose of the workshop}
This \href{https://indico.cern.ch/event/1364999/}{ALEGRO 2024 workshop} was the 5$^{th}$ workshop (\href{http://}{ANAR2017 at CERN}, \href{http://}{Oxford John Adams Institute 2018}, \href{http://}{CERN 2019}, \href{http://}{Hamburg DESY 2023}, \href{https://indico.cern.ch/event/1364999/overview}{Lisbon IST 2024}) organized by the \href{https://www.lpgp.u-psud.fr/icfaana}{ICFA-ANA panel} and a local organizing committee. %
It was endorsed by \href{https://icfa.hep.net/}{ICFA}. %
\newglossaryentry{ALEGRO}{name={ALEGRO},description={Advanced LinEar collider study GROup}} ALEGRO is the Advanced LinEar collider study GROup: an international study group to promote Advanced and Novel Accelerators for High-Energy Physics. 
The purpose of the workshop was to gather the world-wide members of the ANA community interested in developing ANAs for particle and high-energy physics applications. %
It was timely because the report of the US P5~\cite{bib:P5} was released in December 2023, and the roadmap for ANAs, as drawn in Europe as a results of the ESPP process~\cite{bib:ESPP}, was updated since the last workshop~\cite{bib:LDG}, and is being implemented under the leadership of the Laboratory Directors Group (LDG). %
\newglossaryentry{LDG}{name={LDG},description={Laboratory Directors Group}}

\begin{figure}[htb!]
    \centering
    \includegraphics[width=1.0\textwidth]{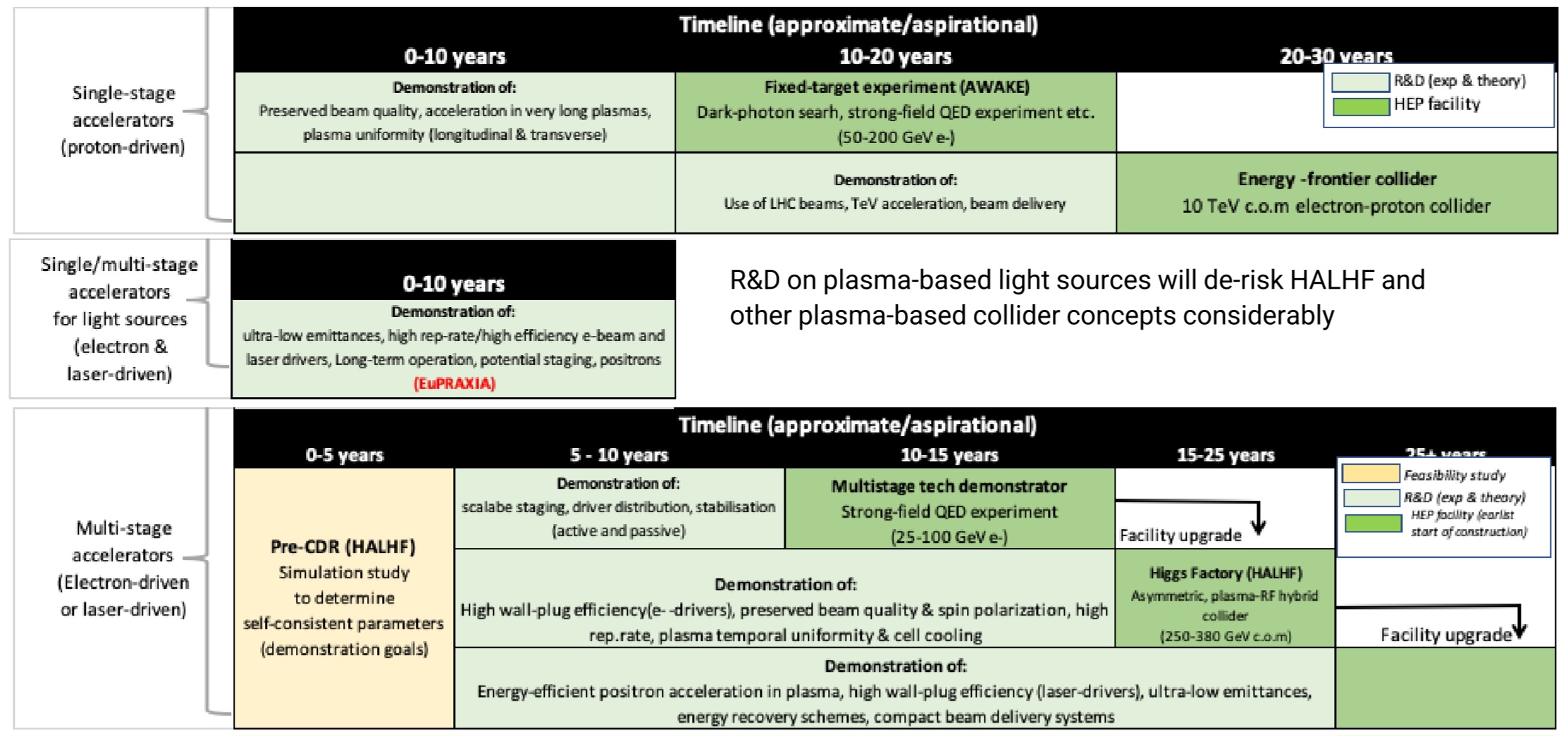}
    \caption{Aspirational roadmap for advanced accelerators as drawn by the LDG. %
    }
    \label{fig:LDGroadmap}
\end{figure}

This updated roadmap (Fig.\ref{fig:LDGroadmap}) includes major efforts for the development of ANAs in Europe led by international collaborations: AWAKE~\cite{bib:AWAKE}, \newglossaryentry{AWAKE}{name={AWAKE},description={Adavanced WAKefield Experiment}} 
EuPRAXIA~\cite{bib:EuPRAXIA} 
\newglossaryentry{EuPRAXIA}{name={EuPRAXIA},description={European Plasma Research Accelerator with eXcellence In Applications}}
and HALHF~\cite{bib:HALHF}. %
\newglossaryentry{HALHF}{name={HALHF},description={Hybrid Asymmetric Linear Higgs Factory}}
It mostly focuses (short- and mid-term) on ANAs applications at the sub-TeV energy scale. %
In contrast, the P5 report calls for the design of a lepton (e$^+$/e$^-$, or muon) collider at the 10\,TeV CM energy or hadron collider at the 100\,TeV pCM energy. %

The workshop program  (\href{https://indico.cern.ch/event/1364999/}{\textsc{https://indico.cern.ch/event/1364999/}}) was organized around two main themes. %
First, to take stock of progress in the ANA field, as pertinent to colliders. %
Second, to get an update on the goals set forward by the ESPP-LDG in Europe and by the P5 in the US, to determine the current state of organization of the community to meet these goals, and possibly propose an active role for ALEGRO, complementing existing organizations. %
Indeed, there are already a number of active collaborations: AWAKE, EuPRAXIA, HALHF in Europe, FACET II in the US. %
\newglossaryentry{FACET II}{name={FACET II},description={Facility for Advanced aCccElerator Tests}}
It was also stated at the workshop that collaboration between national laboratories (SLAC, LBNL, ANL) has already started to address the P5 recommendations. %
\newglossaryentry{PWFA}{name={PWFA},description={Plasma WakeField Accelerator/Acceleration}}
\newglossaryentry{gg}{name={$\gamma\gamma$},description={gamma/gamma or photon/photon (collider)}}

Ample time was scheduled for discussions since the workshop audience gathered representatives of many of the major players in the ANA field. %
A poster session was also part of the workshop. %
Here after we offer a brief summary of the presentations at the workshop. %

\section{Sustainability}
A major topic that was discussed is that of sustainability. %
Indeed, a decisive criterion for the choice for the next collider may well be its sustainability level. %
With smaller footprint and associated reduced need for concrete afforded by the large acceleration gradient, and with lower requirement for power per unit luminosity afforded by the naturally short bunches produced by ANAs, an ANA-based collider may reach a higher sustainability level than any other acceleration schemes. %

The general concepts of sustainability (Denise Völker, DESY, remotely), as well as the particular case of sustainability handling at CERN (Roberto Losito, remotely) were presented and discussed. %
It is particularly important for the ANA community to formulate and communicate well the sustainability aspects of the colliders we are proposing. %
It is thus useful to engage with sustainability experts. %
We note that a new ICFA panel on sustainability was recently formed, highlighting the importance of the topic. %

\section{Projects}
A number of large, international projects exist at different levels of development that explore some accelerator  schemes and  address issues
relevant to application to particle physics or high-energy physics. %

\textbf{AWAKE} (Marlene Turner, CERN) uses the existing, high-energy proton bunches produced by the CERN-SPS to drive wakefields in long plasmas. %
It relies on self-modulation of the long proton bunch ($\sim$cm) to reach acceleration of electron bunches with $\sim$1\,GeV/m (average) gradient. %
It develops scalable plasma sources to reach acceleration length from 10 to 100\,m and use only one long, acceleration stage. %
There is a clear scientific plan to propose in the early 2030's fixed-target experiments for the search for dark photon. %

Short proton bunches, possibly produced by future rapid-cycling synchrotrons, could be used to drive long plasmas to produce electron and positron bunches for a Higgs Factory (Alexander Pukhov). %
A very early proposal was presented. %

The challenge of accelerating positron bunches with collider quality is circumvented by the \textbf{HALHF} collider project. %
Positrons are accelerated in an RF accelerator (31\,GeV). %
Electrons are accelerated in plasma and reach much higher energies (500\,GeV) over the same acceleration distance. %
The collider thus operates in an asymmetric mode. %
The first design can be expanded (Richard d'Arcy) with multiple collision points, energy upgrade, polarized positron beam, even a $\gamma\gamma$ collider option making the project much broader in scope. %

Application to high-energy physics is also included in the \textbf{EuPRAXIA} project (Massimo Ferrario). %
The project addresses issues related to the production of high-quality multi-GeV electron bunches in plasma, able to drive free electron lasers (FELs). %
\newglossaryentry{FEL}{name={FEL},description={Free Electron Laser}}
Accelerators based on laser- and electron-beam-driven plasma wakefields are planned.
The goal is to build two user facilities delivering high-quality electron bunches for various applications, including related accelerator developments, such as positron acceleration, staging, and high-energy physics. %

\textbf{FACET II} has a diverse scientific program and aims at being possibly the only facility where high-energy, high-charge, short e$^+$ bunches and drive e$^-$ bunches may be available (Mark Hogan). %
These are essential to test concepts to accelerate e$^+$ bunches with collider parameters while preserving their quality. %

An indirect contribution to collider studies is a LWFA injector for the large CEPS collider developed in China (Wei Lu, remotely). %
Similar plans exist for PETRA4 and possibly for FCCee. %
\newglossaryentry{LWFA}{name={LWFA},description={Laser WakeField  Accelerator/Acceleration}}
\newglossaryentry{FCCee}{name={FCCee},description={Future Circular Collider e$^-$/e$^-$}}
\newglossaryentry{CEPC}{name={CEPC},description={Circular Electron Positron Collider (China)}}

\section{Positrons and polarized beams}
It is clear that accelerating positron bunches in plasma with parameters and quality required for a collider remains a major challenge for the field (Sebastien Corde). %
While concepts exist to produce test positrons (Gianluca Sarri), the only test facility with positron bunches that would allow for the study of quality acceleration is FACET II (Mark Hogan). %
This heightens the interest in either asymmetric, plasma-based colliders (Richard D'Arcy), or $\gamma\gamma$ and e$^-$/e$^-$ colliders. %
Other possibilities include using structure-based acceleration for the e$^+$ arm of the collider, and only using plasma in the e$^-$ arm of the accelerator. %
The possibility of using proton beams as driver for a Higgs factory was also presented (Alexander Pukhov). %

Because of the short bunches intrinsic to a plasma-based collider, new sources of polarized beams may be required. %
New methods to produce polarized particles for plasma exist (Kristjan P\~oder), but have so far produced long bunches of polarized electrons with low population. %

\section{Numerical simulations}
Numerical simulations are essential to propose, design and understand experiments, and are the first tool for the design of a collider. %
Effective design requires a suite of codes that are adapted to the different parts of the collider, including the appropriate or necessary physics. %
There are a number of initiatives aimed at making transparent to the user the transfer of data from one code to the other. %
One such suite is LASY (Kristjan P\~oder), driven by a multi-institutions collaboration. %
It combines the use of for example FBPIC, Wake T, WarpX and HiPACE. %

A similar initiative is ongoing through BLAST and CAMPA (Alex Huebl), where the emphasis is put in using the proper physics model (2D/3D, full PIC/reduced, low/high resolution, etc.) and codes to best describe each "stage" of a collider, from the injector to the interaction point. %

Determination of global parameters for a plasma-baaed accelerator (PBA) collider have started (Carlo Benedetti). %
\newglossaryentry{PBA}{name={PBA},description={Plasma-Based Accelerator/Acceleration}}
Many options still exist, even for the particular case of laser-driven stages (e.g., guiding method). %
A particular concern is the length of the inter-stage spaces and emittance growth in these spaces. %
All choices and parameters are coupled together (guiding, efficiency, relative energy spread, emittance growth, etc.). %
Therefore, ranking of the importance of the parameters and of the challenges to meet them must be established to make progress. %

Reaching luminosity requires extremely low normalized emittance and transverse size of the bunch in the plasma and wakefields, as well as very tight tolerances. %
Describing the system in numerical simulations thus becomes very challenging. %
Parameters may even reach ranges where usual assumptions, such as mean fields are not valid anymore (Jorge Vieira). %

Including realistic parameters in numerical simulations, such as full laser pulse parameters in the LWFA case, is essential (Kristjan P\~oder) when dealing with requirements for a collider. 

\section{Plasma Sources and Mirrors}

The plasma source is of course central to the PBAs. %
While many types of sources exist (Simon Hooker), the challenges of high-repetition rate and associated cooling, reproducibility, and tunability remain in the concept of a multi-stage collider. %
Long, low-density plasma channels generated by hydrodynamic optical-field-ionization (HOFI) may offer a good option to meet these challenges. %
These have no solid walls, making them immune to damage by laser pulses. %
\newglossaryentry{HOFI}{name={HOFI},description={Hydrodynamic Optical Field-Ionization}}

In the case of laser pulse driving wakefields in the various stages, the driver would be in-coupled using a plasma mirror (Michael Backhouse) placed close to the entrance of the stage. %
The in-coupling of the driver therefore contributes very little to the inter-stage length. %
Various options exist for the 'substrate' to be ionized by the laser pulse from metallic to plastic tape, to liquid jet. %
To minimize emittance growth of the accelerated bunch, the mirror 'substrate' must be as thin as possible to minimize angle scattering, and as close as possible to the beam waist position. %
Debris from the substrate are a major concern when operating at high repetition rate. %
Since the mirror operates at high plasma density, filamentation of the witness bunch with radius larger than the plasma electron skin depth may occur. %
While principle concepts exist, careful studies and engineering design are necessary. %

\section{Structures}
ANA research is dominated by plasma-based concepts because they have demonstrated accelerating gradients much in excess of 1\,GeV/m. %
However, new structures, either metallic or dielectric driven by short electron bunch (Philippe Piot) now operate near the GeV/m. %
In addition, and unlike plasma-based concepts, these are symmetric for electron and positron bunches. %
Therefore, they fit in the general concept of decreasing the length of the accelerator and thus of an electron/positron collider. 
In addition, the two-beam accelerator scheme (TBA, similar to the CLIC scheme) allows for decoupling of the drive and witness beam, therefore offering more possibilities to suppress transverse instabilities. %

The colinear accelerator (CWA) scheme is very similar to PWFA scheme. %

Driving wakefields in a structure with an electron bunch may also allow for reaching higher field in an accelerating cavity or in a gun cavity before breakdown than with a microwave source, because the excitation time is shorter. %

Free electron lasers (FELs) require less stringent beam parameters (emittance, average power, etc.) than collider. %
Therefore developing an FEL driven by an ANA beam can be seen as an intermediate step for an ANA towards a collider. %
Operating FELs are now considering using dielectric structures, either for the whole linac (Argonne) or to boost the energy of the beam emerging from the linac: Swiss FEL, dielectric structure (Evan Ericson), EuPRAXIA@SPARC\_LAB: plasma (Massimo Ferrario). %
For example a slab dielectric structure could be used as dechirper. %

Dielectric structures (slab or others) can be used to preserve beam quality (Oznur Apsimon). %

\section{Beam delivery at interaction point}
Discussions led to the conclusion that, to strengthen it case, ALEGRO must engage the particle physics community that would use the advanced collider. %
ALEGRO must engage in developing self-consistently the beam delivery system and detector(s) for the proposed collider. %
Preliminary considerations show that the beam delivery system developed for CLIC~\cite{bib:CLIC} has many features also suitable for an ANA-based collider (Vera Cilento). %
However, there is a great premium in having low relative energy spread and reproducible energy from bunch to bunch. %
A significant length of the beam delivery system consist of collimators that protect the detectors from off trajectory/momentum events and their length cannot be reduced. %

The plasma community has started including in numerical simulation codes used to describe the acceleration processes in plasma, the tools to be able to describe the ''plasma physics'' at the interaction point, including radiation and QED effects (Thomas Grismayer).  %
\newglossaryentry{QED}{name={QED},description={Quantum Electro-Dynamics}}
This is an important progress toward a self-consistent, and possibly higher level of sustainability design, for example because of possible higher luminosity to power ratio. %

While colliders usually use ''flat beam'', plasma acceleration favors symmetric ''round beams''. %
Numerical simulation results relevant for HALHF (Severin Diedrich), show that in plasma coupling between the two emittance planes occurs, which strongly increases the vertical (low) emittance. %
While solutions may exist to mitigate the effect over the first few stages of an accelerator, the issue may re-surface in later stages. %
Round beams may also reduce emission of betatron radiation, which may be a limitation in energy spread and emittance at multi-TeV energies. %
Differences between ''flat'' and ''round'' beams must also be taken into account for the plasma physics at the interaction point (Thomas Grismayer). %
The theoretical understanding of beam disruption helps in designing better beam dynamics simulations and predicting the outcomes of collisions in linear colliders. Managing disruption is critical and involves a combination of advanced particle beam physics, strong field QED, and plasma physics to optimize both the beam quality and the interaction point environment. %

\section{Other Topics}
In- and out-coupling of the driver is a challenge for a co-linear accelerator. %
Curved plasma channels could be considered to replace plasma mirrors (laser driver) (Boyuan Li) or magnetic optics (particle driver). %
However, experimental demonstrations with intense pulses and bunches suitable to drive meter scale plasma stages are required.  

Compact hybrid LWFA-PWFA accelerators may provide opportunities to test some of concepts that could be applicable to a collider though with very low energy beams (Susanne Schroebel). 

\section{Role of ALEGRO}
ALEGRO sees its role in fostering progress towards an ANA-based linear collider.
This role is in line with the P5 recommendation and with the ambitions of the European community. %
Concepts or straw-person designs were put forward for an ANA-based electron-positron collider. %
ALEGRO has submitted a contribution for the ESPP process~\cite{ALEGROESPP}. %
However, ALEGRO currently has no source of funding to support any work. %

In the US, funding for ANA R\&D predominantly originates for the HEP branch of the US-DoE. %
Much of the development occurs at national laboratories (e.g., Lawrence Berkeley, SLAC, Argonne, and Brookhaven National Laboratories)  with very important contributions for a number of universities. %
In Europe, funding is provided by national agencies (CNRS-France, BMBF-Germany, etc.) for many university groups and R\&D also occurs at national (DESY-Germany, CLF-UK, INFN-Italy) or international (CERN, ELI) scale laboratories. %

Therefore, ALEGRO has the ambition to apply for funding, in particular in Europe, but possibly in the US through US/EU NSF funding opportunities. 
The purpose is to support work towards a 10\,TeV collider between the European, the US, and the Asian groups. %
At this time, funding can only cover numerical simulation studies. %

\newglossaryentry{CDR}{name={CDR},description={Conceptual Design Report}}
\newglossaryentry{LPA}{name={LPA},description={Laser Plasma Accelerator}}
\newglossaryentry{SWFA}{name={SWFA},description={Structure WakeField Accelerator/Acceleration}}
\newglossaryentry{DoE}{name={DoE},description={Department of Energy (US)}}

\newpage

\chapterimage{images_head_sandwhitep}
\chapter{Contributions}

We include "one-page" summaries of the various contribution as provided by the speakers. %


\section{R\&D Roadmap of the European Particle Physics Strategy: Update on Plasma Accelerator R\&D \\ \small{\textit{W. Leemans\textsuperscript{1}, R. Pattathil\textsuperscript{2}}} \normalsize}

\textit{\textsuperscript{1}Accelerator Division, DESY, Hamburg, Germany}\\
\textit{\textsuperscript{2}Central Laser Facility, STFC Rutherford Appleton Laboratory, UKRI, United Kingdom}
\\
\\ 
The European Strategy for Particle Physics Strategy published in 2022~\cite{Leemans_1} outlines a roadmap for European accelerator R\&D for the next five to ten years, covering five topical areas including investigations of the potential of laser/plasma acceleration. The implementation of activities in this roadmap will provide the evidence base to support subsequent decisions on prioritisation, resourcing and implementation. This report provides an update on these activities. 

The development of a concept for a plasma-based collider is a multi-decadal activity, advancing all aspects of beam-driven and laser-driven plasma accelerators and associated technologies with the aim of building such a device in 20 years from now. Although we envisage to have a CDR by the end of the decade, in the interim, we plan to have a pre-CDR for a plasma-based collider by 2026 as well as the demonstration of several key aspects in plasma wakefield acceleration, which are relevant for a plasma-based colliders. The programme is divided into workpackages spanning four major themes: (a) Plasma Accelerator Feasibility studies via theory and modelling, (b) Development of ancillary technologies for high repetition rate plasma accelerators (c) Experimental studies towards solving some key R\&D challenges using existing laser/electron/proton-driven plasma accelerators and (d) Interim demonstrations of particle physics-relevant experiments. 

It is agreed that the Pre-CDR will be based on the HALHF concept (Hybrid Asymmetic Linear Higgs Factory), an asymmetric energy e$^+$/e$^-$ collider design with a plasma-accelerator-based 500\,GeV electron arm and a 31\,GeV positron arm driven by a conventional linac~\cite{Leemans_2}. If successful, this would allow a Higgs factory to be built at $\sim$1/4 of the cost of a fully conventional collider, with a potentially lower carbon footprint because of the reduction in size. In the first generation, the plasma accelerator arm of HALHF will be based on particle-driven wakefield accelerator (PWFA) because of their higher Technical Readiness Level (TRL) (4-5) but a laser-driven plasma accelerator (LPA) arm (currently at TRL – 3) could be incorporated for future upgrades once the TRLs improve. The modelling activities focusing on start-to-end simulations are progressing very well, aided by the recent ERC grant by Carl Carl Lindstrøm.

The successful implementation and future proofing of HALHF-based depends on addressing some key R\&D challenges for plasma accelerators as identified in the work packages, ranging from identifying ways to accelerate positrons, preserving spin and polarisation to developing high-repetition rate laser drivers and targetry. Substantial progress has already been made in some of them in the last year (e.g. demonstration of 10\,m long discharge plasma source prototype in AWAKE, understanding recovery timescales in plasma at FLASHForward etc.), some others (e.g., staging of multiple plasma accelerators) are due to be tested in the coming months. %
\vspace{1cm}
\section{Physics considerations for laser-plasma linear colliders \\
\small{\textit{C. Benedetti, C.B. Schroder, D. Terzani, S. S. Bulanov, J. Osterhoff, E. Esarey, and C.G.R. Geddes}}\normalsize}

\textit{BELLA Center, Lawrence Berkeley National Laboratory, Berkeley, CA, USA}
\\
\\
Laser-plasma accelerators LPAs~\cite{benedetti_1} are capable of sustaining accelerating gradients of 10 - 100\,GV/m, 100 - 1000 times that of conventional radio-frequency-based technology, and, hence, are regarded as potential drivers for a future, less-expensive, multi-TeV e$^+$/e$^-$ (or $\gamma\gamma$) collider. A plasma-based collider will be based on staging independently-powered LPAs, where the drivers powering each stage are coupled in and out of each LPA stage over a short distance by means of plasma mirrors. Staging technology provides a path to reach high particle beam energies while keeping a high average accelerating gradient without increasing the laser driver energy. Also, LPAs intrinsically accelerate short particle bunches, much shorter than that of conventional technology, which leads to reductions in beamstrahlung and, hence, savings in the overall power consumption to reach a desired luminosity~\cite{benedetti_2}.

A conceptual design of an LPA-based linear collider has not been completed. However, several preliminary studies have been performed~\cite{benedetti_3, benedetti_4}. Such studies have been instrumental in helping to identify the laser and plasma parameters needed to guide the R\&D effort towards a collider. For instance, the properties of the LPAs depend on the choice of the operational plasma density, which sets the energy gain provided by the LPA stage, its length, the characteristic charge that can be accelerated, the laser energy required to drive the stage, etc. Considerations concerning minimization of the total accelerator length, wall-plug power, and beamstrahlung effects, while reaching the desired luminosity (e.g., $>10^{34}\textrm{cm}^{-2}s^{-1}$ for $>1$\,TeV center of mass energies) suggest operating at a density on the order of $10^{17}\,\textrm{cm}^{-3}$. This requires laser drivers with an energy of 10s of Joules operating with a repetition rate of 10s of kHz. Each LPA stage will then provide multi-GeV energy gains to beams with a charge of 100s of pC, over a distance of tens of cm. Reaching TeV-class energies necessitates cascading 100s of stages. High efficiencies (several 10s of percent) from wall-plug to laser driver production, from driver to wake excitation, and from wake to beam are required. Finally, in order to ensure the focusability of the particle beam at the interaction point and enable staging, high beam quality (i.e., relative energy spreads < 1\% and emittances <100\,nm) must be maintained throughout the accelerator. Design of LPA stages providing for a fixed laser technology (i.e., fixed laser energy and wavelength) high-quality, high-energy, and high-efficiency acceleration of electrons and positrons beams is underway \cite{benedetti_5}.  Based on these stages, high-level collider parameters for a 1, 3, and 15\,TeV electron-positron collider, and for a 15\,TeV $\gamma\gamma$ collider have been proposed~\cite{benedetti_6}.

A critical next step towards the development of a multi-TeV, LPA-based collider is the delivery of an end-to-end design concept, including cost scales, with self-consistent parameters throughout. Significant research is required to develop both the laser-plasma accelerator technology and the auxiliary systems compatible with laser-plasma-accelerated particle beams. A summary of the open problems, possible solutions, the status of the current R\&D effort, and the roadmap towards a plasma-based collider are discussed in Ref.~\cite{benedetti_6}. 

This work was supported by the Director, Office of Science, Office of High Energy Physics, of the U.S. Department of Energy under Contract No. DE-AC02-05CH11231.
\section{Advances in SWFA R\&D for integration in linear colliders\\
\small{\textit{P. Piot, and J. G. Power}}\normalsize}

\textit{Argonne National Laboratory, Argonne, Illinois 60439, USA}
\\
\\
The concept of structure wakefield acceleration (SWFA) offers a pathway to GV/m accelerating fields with operating frequencies ranging from X-band to THz. The implementation of SWFA can be based on collinear wakefield acceleration (CWA) or two-beam acceleration (TBA). CWA shares many features of the PWFA concept where a drive beam produced high-accelerating field experiences by a main bunch located propagating along the same path with the proper delay. TBA is unique to SWFA and uses a drive beam to excite high-power electromagnetic (EM) pulses in a power-extraction structure which are outcoupled and transferred to an accelerating structure used to boost the main-beam energy. TBA has the advantage of decoupling the beam dynamics of the drive and main beams at the expense of having two parallel beamlines with independent focusing, diagnostics, and other auxiliary components.   

To date, the SWFA community has primarily concentrated on a Linear Collider (LC) straw-person design utilizing the TBA implementation due to its maturity. A concept akin to CLIC design scaled to 26\,GHz was proposed in~\cite{Gai}. Meeting LC requirements entails producing an EM pulse peak power of 1\,GW in the power-extracting structure. While peak power close to 0.6\,GW~\cite{Picard} was achieved over the past three years, it is currently constrained by the quality and energy of available drive beams. Additionally, the staging of TBA was successfully demonstrated using X-band structures~\cite{Jing}. Recently, surface fields approaching 0.6\,GV/m were reached in a photoinjector powered by the TBA scheme~\cite{Tan}. The ability of TBA to generate high-peak-power short EM pulses, with durations less than 10\,ns at X-band frequencies, is key to achieving these high-peak fields. The short EM pulses significantly reduce breakdown rates and enable higher yields compared to klystron-powered structures. Comparable sub-GV/m peak fields were also obtained in other accelerating structures~\cite{Shao}. Over the next five years, a comprehensive TBA concept aimed at demonstrating the generation and acceleration of bright beams to 0.5\,GeV is planned, potentially driving a free-electron laser operating in the vacuum- or extreme-ultraviolet (V/EUV) regime~\cite{Piot}. 

The exploration of the Collinear Wakefield Acceleration (CWA) option for SWFA is also underway. Thus far, an integrated design of a multi-user FEL facility was proposed, which motivated the investigation of a single-stage CWA concept~\cite{Zholents}. Moreover, extensive research has been conducted on the topic of drive-beam shaping, resulting in the demonstration of a record transformer ratio of $\sim$5~\cite{Gao}. Finally, efforts to study the transverse beam stability and mitigate the beam-breakup instability are actively underway, both theoretically~\cite{Baturin} and experimentally~\cite{Lynn,Lynn2}. An experimental program focused on developing an integrated CWA-based accelerating module capable of doubling the energy of the AWA accelerator in Argonne is currently in the planning phase. It should be also stressed that CWA implementation of SWFA shares many commonalities with PWFA, thereby offering opportunity for collaboration on the topics of shaped drive beam generation, and phase space manipulation and cooling of the electron and positron main bunches.   

Both the CWA and TBA concepts ultimately depend on the development of efficient numerical tools. A significant aspect of SWFA modeling involves designing structures with optimal RF performances and geometry to control the topology of the electromagnetic field. However, integrated simulation of the beam dynamics, particularly in CWA, is still lacking, with only a handful of particle-in-cell finite-difference time-domain computer programs incorporating electromagnetic boundaries and macroscopic properties relevant to SWFA (WARP-X~\cite{Tan2} being the exception). 

\section{Sustainability\\
\small{\textit{D. Völker\textsuperscript{1}, and R. Losito\textsuperscript{2}}}\normalsize}

\textit{\textsuperscript{1}DESY, Hamburg, Germany}\\
\textit{\textsuperscript{2}CERN, Geneva, Switzerland}
\\
\\
\begin{itemize}
    \item The last few years have shown that the challenges facing our society continue to increase and require a solution much more quickly.
    \item Scientific facts on Climate Change: Research Centers as scientific institutions have to act accordingly! 
    \item The SDGs, the Paris Agreements and many more give as a clear path ahead .
    \item In view of the societal challenges, the science centers have two responsibilities. They enable research for a sustainable society, and at the same time great attention must be paid to operating the large devices sustainably.
    \item Research infrastructures enable basic research in order to be able to make the world more sustainable and thus contribute to solving the urgent questions of our time.
    \item At the same time, the research infrastructures are working flat out to become more and more sustainable themselves.
    \item It all starts with the right design of machines and buildings.
    \item Most important fields of action: 
    \begin{itemize}
        \item Large scale civil construction $\rightarrow$ innovative sustainable construction
        \item High resource consumption in operation $\rightarrow$ Focus on energy efficiency and critical materials 
        \item Specific technologies needed $\rightarrow$ innovation, R\&D, cooperation (that in turn can be transferred back into economy) 
        \item Big data $\rightarrow$ Green IT and data reduction concepts 
        \item Environmental reporting (GRI)
        \item Energy management certification (DIN ISO 50001)
        \item Materiality analysis 
    \end{itemize}
    \item Project examples were given: 
    \begin{itemize}
        \item Waste heat usage
        \item Civil construction and building greening
        \item Permanent magnets
        \item Tunnel Temperature
        \item SF6 and other Green House Gases used for particle detection
        \item Responsible Procurement
        \item Life Cycle Assessment (LCA) for CLIC \& ILC
        \item Concrete for tunnel construction
        \item HTS
        \item Superconductivity RF 
    \end{itemize}
The presentations were followed by a discussion around: 
    \begin{itemize} 
        \item What to assess in an LCA or a footprint $\rightarrow$ LDG working group
        \item Nuclear power pro and con
        \item Magnets and decommissioning
        \item Networking/exchange on technical and process questions
        \item Possible resource demand reductions via plasma
        \item Data centers and data reduction
        \item Need for a cultural change 
    \end{itemize}
\end{itemize}
The main conclusion was that everybody can and should do what is in her/his power to make the machines more sustainable.  
\section{Prospects and challenges for high-repetition-rate plasma sources for future colliders \\ 
\small{\textit{S. Hooker, and R. D'Arcy}}\normalsize}
\noindent \textit{University of Oxford, United Kingdom}
\\
\\
A collider based on plasma accelerators would have to operate at a high mean pulse repetition rate in order to achieve the required luminosity, which places stringent demands on the plasma source at the heart of each plasma stage. Here we outline those demands  and provide a short summary of some efforts to meet these.

It is useful first to estimate the energy stored in, and hence cooling requirement of, an operating plasma accelerator stage. The energy density of the plasma wakefield is of order $(1/2) \epsilon_0 E_0^2$, where $E_0$ is  the peak electric field within the wakefield. A wakefield with $E_0 = \SI{10}{GV.m^{-1}}$ has an energy density of approximately \SI{400}{J.cm^{-3}}, corresponding to $\sim \SI{3}{J}$ for a \SI{1}{m} long, \SI{100}{\micro m} diameter plasma. Even assuming an optimistic value of $\eta = 75\%$ for the efficiency with which this energy is transferred to a witness bunch, one concludes that after acceleration an energy of order \SI{1}{J} will remain in each metre of plasma. It is easy to show that this energy is 3 or 4 orders of magnitude larger than that required to produce the plasma itself.
 
A similar estimate can be reached by considering the required cooling per unit length of a plasma accelerator, which is given by, $d P_\mathrm{cool} / d z = E_0 q_b f_\mathrm{rep} [\eta^{-1} - 1]$, where $q_b$ is the bunch charge and $f_\mathrm{rep}$ is the bunch repetition rate. Assuming values \cite{Schroeder.2010, Schroeder.2023} of $E_0 =\SI{3.3}{GV.m^{-1}}$, $q_b = \SI{0.2}{nC}$, $f_\mathrm{rep} = \SI{50}{kHz}$, and $\eta = 75\%$, we find a required cooling gradient of \SI{11}{kW.m^{-1}}. This cooling gradient is comparable to that required by the CLIC design.

In the light of these estimates, one can envisage operating a plasma accelerator at high repetition rates by one of several strategies. First,  the plasma could be replaced by fresh plasma for each witness bunch. Assuming that plasma could be moved at approximately the speed of sound ($c_s \approx \SI{1.3}{km.s^{-1}}$ for hydrogen gas), then the time to move the plasma transversely a distance $d$ is $\Delta t = d/c_s$. For $d = \SI{1}{mm}$ and $c_s \approx \SI{1.3}{km.s^{-1}}$, then $\Delta t \approx \SI{1}{\micro.s}$, from which we conclude that $f_\mathrm{rep} \sim \SI{1}{MHz}$ might be possible.

A second strategy is to wait for the plasma to recover. The unused wakefield energy is expected to drive a complex chain of processes that includes wakefield decay, ion motion, ionization of the surrounding gas, heat conduction, and plasma motion and expulsion. Zgadzaj et al. \cite{Zgadzaj.2020} have investigated these processes over a $\sim \SI{1}{ns}$ time-scale following wake excitation by the \SI{20}{GeV} SLAC linac. However, the flow of energy over longer timescales has yet to be investigated. D'Arcy et al. \cite{D’Arcy.2022} have shown experimentally that the plasma of a PWFA can recover in less than \SI{100}{ns}, indicating the potential for $f_\mathrm{rep} \lesssim \SI{10}{MHz}$, although these measurements do not include mean-power effects.

A third strategy is to use the same plasma for multiple acceleration cycles before replacing it. This might be particularly appropriate to accelerators driven in capillary structures, such as capillary discharge waveguides. For these systems the timescale for ejection of the plasma is long \cite{Garland.20211qf}, $\sim \SI{5}{\micro.s}$, compared to the plasma recovery time $\lesssim \SI{100}{ns}$, and hence it might be possible to accelerate a train of many bunches before the plasma needs to be replaced.

A fourth strategy would be to remove unused wakefield energy with one or more additional laser pulses. In principle this can be achieved by an out-of-phase laser pulse located after the witness bunch. This ``recovery'' pulse will drive its own, out-of-phase wakefield that destructively interferes with the primary wake, leaving behind a wakefield with greatly reduced amplitude. The energy of the primary wakefield is removed from the plasma by spectral blue-shifting of the trailing laser pulse, and in principle this energy can be recovered. There has been little work today on energy recovery in plasma accelerators, but preliminary work by Cowley et al. \cite{Cowley.2017} showed that the relative amplitude of a linear wakefield could be reduced by a recovery laser pulse from 0.6\% to 0.34\%, corresponding to removal of 70\% of the original wake energy. It would be interesting, therefore, to explore this possibility further.

The considerations above demonstrate that operating a plasma accelerator at high repetition rates would place severe demands on the plasma source. Here we summarize very briefly some work aimed at meeting these requirements.

The AWAKE project has formed a collaboration of five institutes (Max Planck Institute for Plasma Physics, EPFL, University Wisconsin-Madison, IST Lisbon, and Imperial College London), with a dedicated source laboratory at CERN, to develop plasma sources for future AWAKE runs. These sources include a \SI{10}{Hz} helicon plasma source, and a direct current cold-cathode discharge source that has already generated \SI{10}{m} long plasmas.

In SPARC LAB, INFN work is being undertaken on the development of multi-Hz capillary discharges. This work includes testing of different capillary materials, and development of high-repetition-rate discharges circuits. This group has demonstrated operation of a capillary discharge at $f_\mathrm{rep} = \SI{50}{Hz}$.

A programme to develop high-repetition-rate capillary discharges is being undertaken by a collaboration between DESY and Oxford. This work includes development of 1) capillary discharges with a MHz burst mode regime of operation (DESY) and 2) actively cooled and highly robust capillaries capable of sustained 10 kW/m operation (Oxford).

The group at Oxford is developing all-optical plasma channels, which are of particular interest for laser-driven plasma accelerators operating at high pulse repetition rates. These hydrodynamic optical-field-ionized (HOFI) plasma channels \cite{Shalloo.2018} have been demonstrated to guide high-intensity ($\sim \SI{E18}{W.cm^{-2}}$) laser pulses through \SI{100}{mm} long channels of axial density $\sim \SI{E17}{cm^{-3}}$ with low losses \cite{Picksley.2020}. They have also been operated at a repetition rate of \SI{0.4}{kHz} for \SI{6.5}{hours} with no degradation of the channel properties \cite{Alejo.2022}.
\section{Physics considerations for laser-plasma linear colliders \\
\small{\textit{M. Backhouse, and Z. Najmudin}}\normalsize}

\textit{The John Adams Institute for Accelerator Science, Imperial College, London SW7 2BZ, United Kingdom}
\\
\\
Achieving high electron energies with plasma accelerators will likely require multiple independently driven stages. 
To be useful for frontier particle physics applications, such a multi-stage accelerator must achieve high luminosities. In addition, since the potential for cost reduction that motivates LWFA development comes primarily from the large accelerating gradients that can be reached in plasmas, maximising the average accelerating gradient of the entire system is vital. Coupling new drive laser pulses into LWFA stages without increasing the overall length of the accelerator may be done using thin-film plasma mirrors, which rely on a flat, overdense plasma to reflect the laser pulse close to its focus. The slow focussing that produces optimal laser spots for LWFA can be done orthogonally to the electron beam axis, minimising the inter-stage distance. 

Plasma mirrors increase the beam emittance, which will reduce the ultimate luminosity.  Elastic scattering in the plasma mirror increases the transverse momentum spread in the beam, resulting in an uncorrelated normalised emittance increase. The strength of the scattering scales with the density of the mirror and its thickness. The effect on emittance is much greater if the beam is allowed to drift before undergoing scattering; to mitigate this drift effect the electron beam should not double in diameter before scattering. 

The laser can drive currents in the plasma mirror that may support $\sim$\SI{10}{\kilo\tesla} strength magnetic fields~\cite{backhouse:Raj2020}. At laser intensities relevant to LWFA, sub-micron scale fields can saturate in strength before the electron beam arrives. A divergence increase results from the spatially dependent deflection of the electrons, which increases the emittance. The magnitude of the divergence increase is determined by the field strength, $B\propto a_0 n_h$, where $a_0$ and $n_h$ are the normalised laser vector potential and hot electron density respectively. 

At the end of the accelerating stage, the remaining laser energy must be extracted using a plasma mirror. The electron beam can undergo inverse Compton scattering as it passes through the tail of the laser pulse, resulting in spectral broadening~\cite{backhouse:Streeter}. If unmitigated, this effect 
may cause beam loss, $\sim$10\% at 20 GeV. Reducing the laser intensity incident on the mirror, using shaped plasma mirrors, incomplete dephasing, and the use of later phases of the plasma wave to accelerate the beam are proposed mitigation methods.

The plasma mirror must be positioned with sub-micron accuracy to limit energy jitter to less than 1\% after 100 stages. This ensures that the delay between the electron beam and the laser is constant. To support kHz repetition rates a tape-based mirror must spool at approximately \SI{1}{\meter\per\second}, and this operation must be stable over hours of operation. Achieving these objectives simultaneously presents a technical challenge. A tape drive was developed capable of producing \SI{0.7}{\micro\meter} rms positional jitter while spooling Kapton tape at an average of \SI{0.5}{\meter\per\second}, and was shown to be stable during long operation, with low positional drift over 8 hours of operation~\cite{backhouse:Xu2023}. Future plasma mirror replacement devices must consider the effect of ablating solid material at high repetition rates on long-term operation under vacuum, in which case ultra-thin film or liquid targets may be preferable.

\section{LASY: LAser manipulations made eaSY \\ 
\small{\textit{M. Th\'evenet\textsuperscript{1}, I. A. Andriyash\textsuperscript{2}, L. Fedeli\textsuperscript{3}, A Ferran Pousa\textsuperscript{1}, A. Huebl\textsuperscript{4}, S. Jalas\textsuperscript{1}, M. Kirchen\textsuperscript{1}, R. Lehe\textsuperscript{4}, K. P\~oder\textsuperscript{1} (presenter), R. J. Shalloo\textsuperscript{1}, A. Sinn\textsuperscript{1,5}, and J.-L. Vay\textsuperscript{4}}}\normalsize}

\textit{\textsuperscript{1}Deutsches Elektronen-Synchrotron DESY, Notkestr. 85, 22607 Hamburg, Germany} \\
\textit{\textsuperscript{2}Laboratoire d’Optique Appliquée, ENSTA Paris, CNRS, Ecole Polytechnique, Institut Polytechnique de Paris, 828 Bd des Maréchaux, 91762 Palaiseau, France} \\
\textit{\textsuperscript{3}Université Paris-Saclay, CEA, LIDYL, 91191 Gif-sur-Yvette, France} \\
\textit{\textsuperscript{4}Lawrence Berkeley National Laboratory LBNL, Berkeley, California 94720, USA} \\
\textit{\textsuperscript{5}Universit{\"a}t Hamburg UHH, Mittelweg 177, 20148 Hamburg, Germany}
\\
\\
While multiple works demonstrated the importance of using realistic laser profiles for simulations of laser-plasma accelerators to accurately reproduce experimental measurements, the handshake between experiments and simulations can be challenging. Similarly, transferring a laser pulse from one code to another, as needed for start-to-end simulations, may require some error-prone manipulations. In this contribution, we presented LASY (LAser manipulations made eaSY), a new open-source Python library to simplify these workflows. Its main features are:
\vspace{0.5cm}
\begin{itemize}
    \item{\textbf{Read various input sources}}
    \begin{itemize}
        \item Experimental measurement: transverse intensity profile, spectrum, 3D profiles e.g. INSIGHT [\href{https://doi.org/10.1364/OE.26.026444}{A Borot and F Quéré (2018) Optics Express 26 26444–61}].
        \item Simulation output: envelope/full field on a regular grid, 3D cartesian or 2D cylindrical with azimuthal decomposition...
        \item Analytic profiles: Gaussian, Hermite-Gauss, Laguerre-Gauss…
    \end{itemize}
    \vspace{0.5cm}
    \item{\textbf{Perform operations on laser profiles}}
    \begin{itemize}
        \item Conversions: electric field/vector potential, field/envelope, time/space, 2D cylindrical/3D cartesian.
        \item Smoothing/filtering.
        \item Vacuum propagation (powered by Axiprop [\href{https://github.com/hightower8083/axiprop}{https://github.com/hightower8083/axiprop}]).
    \end{itemize}
    \vspace{0.5cm}
    \item{\textbf{Write to file complying with a standard.} The LaserEnvelope extension of the openPMD standard was developed for this purpose. The file can then be used for further postprocessing, or as an input for a numerical simulation. }
\end{itemize}

\vspace{0.5cm}

\begin{figure}[htb!]
    \centering
    \includegraphics[width=0.75\columnwidth]{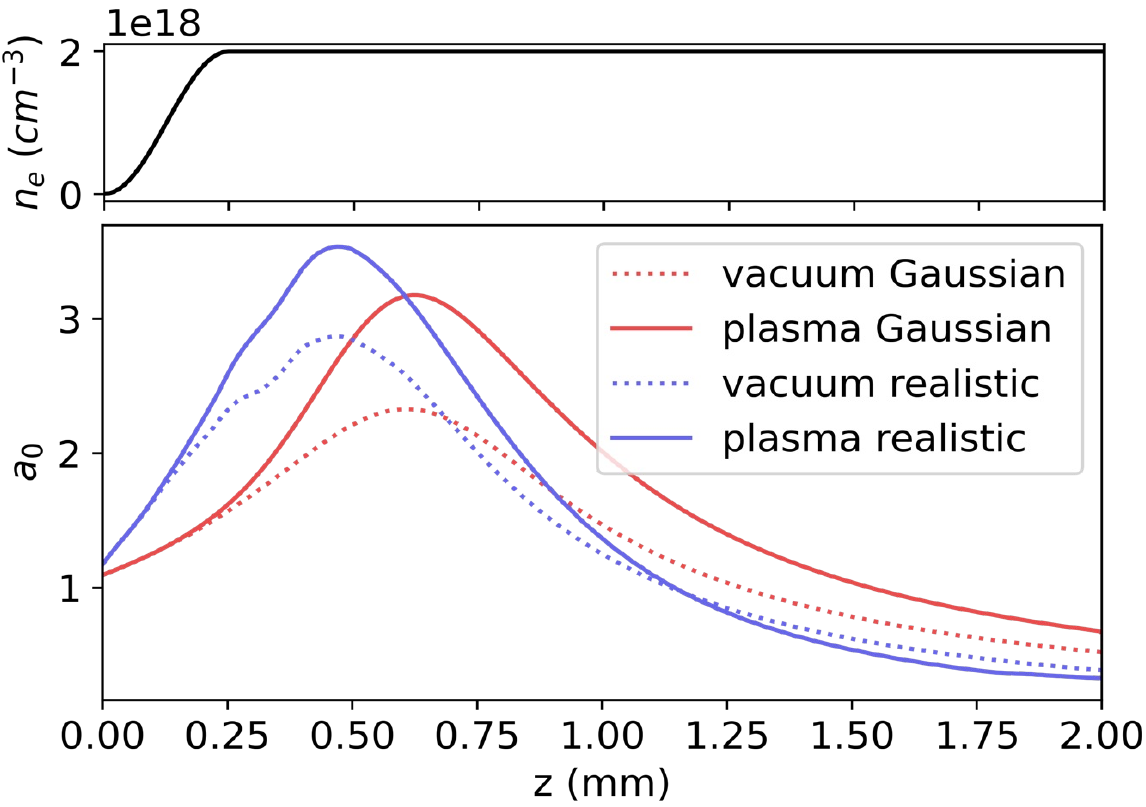}
    \caption{Demonstration of the experimental data cleaning built into LASY. Propagation of a measured profile in a plasma and in vacuum, compared with an equivalent Gaussian profile.}
    \label{fig:figure1}
\end{figure}

\begin{figure}[htb!]
    \centering
    \includegraphics[width=0.75\columnwidth]{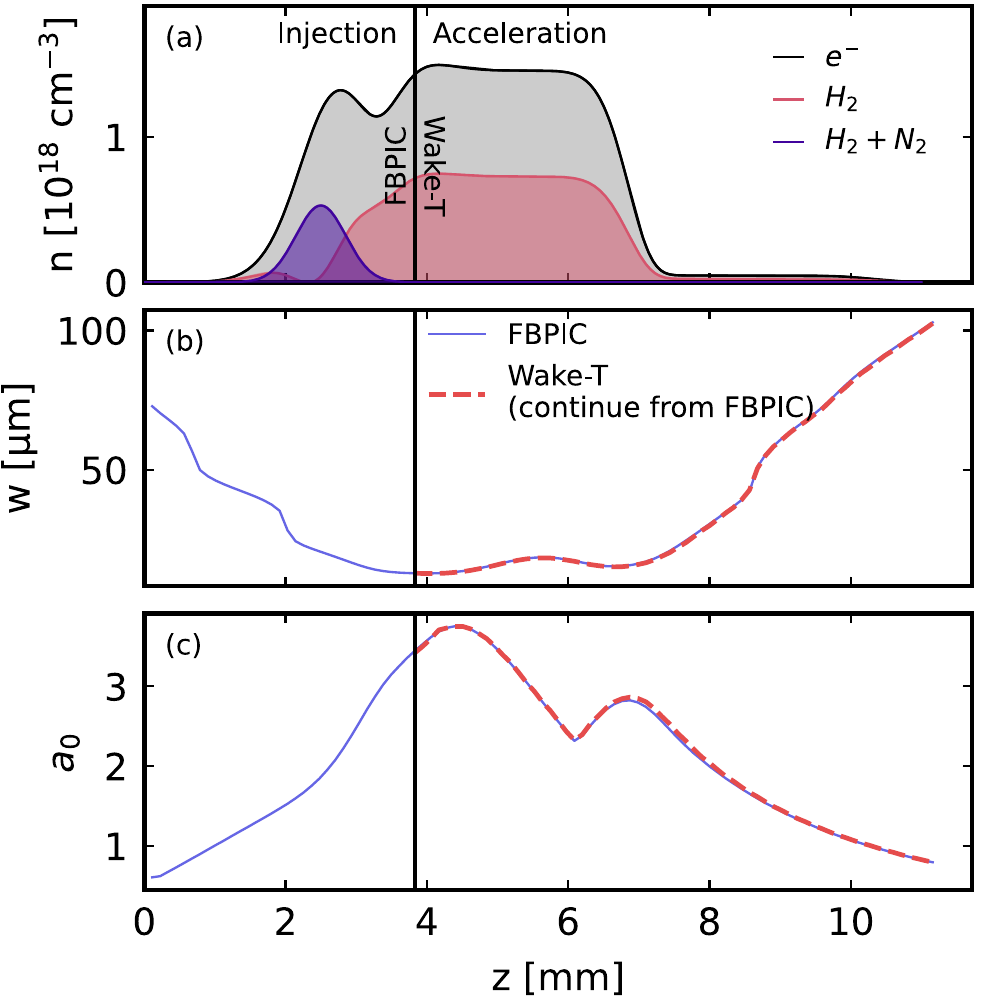}
    \caption{Laser wakefield acceleration simulation using FBPIC for beam injection and Wake-T for the subsequent acceleration, using LASY to convert and transfer the laser profile between the two codes at z = 3.9 mm. (a) plasma profile. (b) and (c) show the laser pulse width and peak normalized amplitude, respectively, comparing the workflow with both codes to a full FBPIC simulation.}
    \label{fig:figure2}
\end{figure}

The capabilities of LASY are illustrated on typical workflows: run a 3D simulation from an experimentally measured laser profile (\ref{fig:figure1}); and transfer the output of an electromagnetic particle-in-cell simulation to a quasistatic one (\ref{fig:figure2}).

Overall, LASY is a versatile library to simplify the use of realistic profiles for higher-fidelity simulations and start-to-end workflows. It adopts community standards to encourage multi-code pipelines for higher efficiency. LASY is developed as part of an international collaboration, with the goal to combine efforts to expose more features (optics, phase manipulations, spatio-temporal coupling, propagators, filters, analytical profiles, readers, etc.) to the community in a user-friendly manner. For more information:
\vspace{0.5cm}
\begin{itemize}
\item Preprint (with all relevant references) \href{https://arxiv.org/abs/2403.12191}{https://arxiv.org/abs/2403.12191}
\item Repository \href{https://github.com/LASY-org/lasy}{https://github.com/LASY-org/lasy} 
\item Documentation \href{https://lasydoc.readthedocs.io}{https://lasydoc.readthedocs.io}
\end{itemize}

\vspace{1cm}

\section{Multistage LWFA based on curved plasma channels\\
\small{\textbf{\textit{\textit{Boyuan Li\textsuperscript{1}, Xinzhe Zhu\textsuperscript{1}, Min Chen\textsuperscript{1}, Zhengming Sheng\textsuperscript{1,2}, Jie Zhang\textsuperscript{1,2}}}}}\normalsize}

\begin{figure}[htb!]
    \centering
    \includegraphics[width=1.0\textwidth]{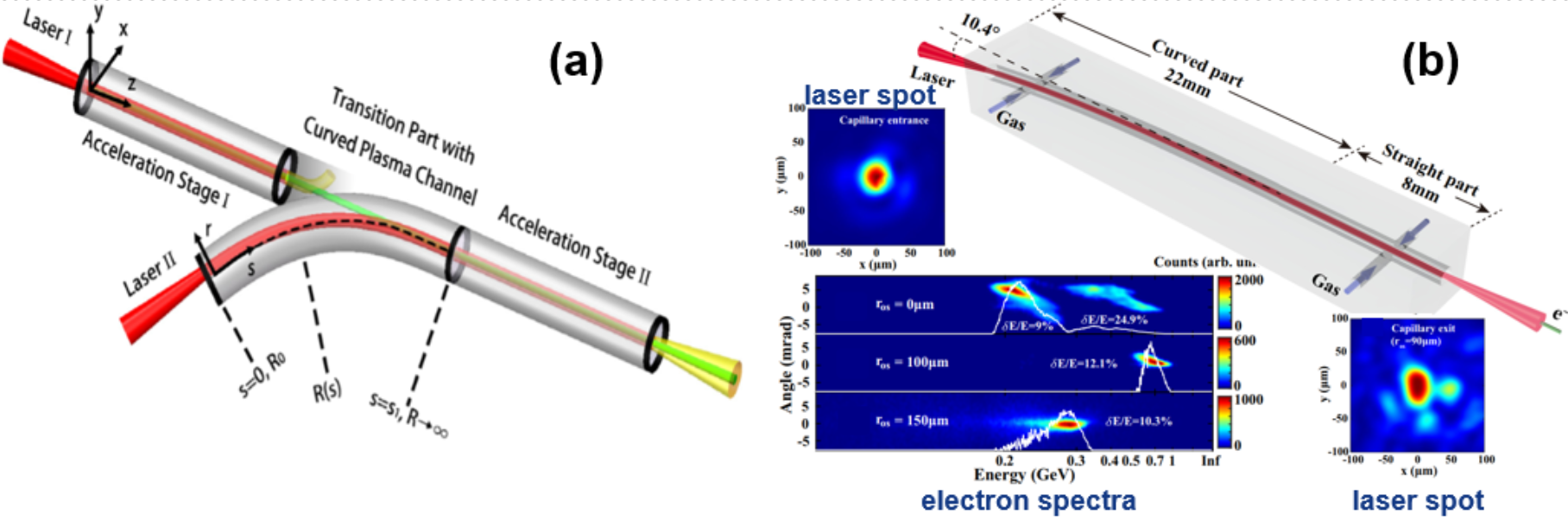}
    \caption{(a) Scheme of multistage LWFA based on curved plasma channels. (b) Experimental demonstration of laser guiding and wakefield acceleration in a curved plasma channel.  %
    }
    \label{fig:schematic}
\end{figure}

\textit{\textsuperscript{1}Key Laboratory for Laser Plasmas, School of Physics and Astronomy, Shanghai Jiao Tong University, Shanghai 200240, China}
\textit{\textsuperscript{2}Tsung-Dao Lee Institute, Shanghai Jiao Tong University, Shanghai 200240, China}
\\
\\
Laser wakefield accelerators (LWFA) have attracted considerable attention as a promising new accelerator technology. They are capable of supporting enormous acceleration gradients, as high as hundreds of GeV/m. Multistage coupling of laser wakefield accelerators is essential to overcome laser energy depletion for high-energy applications such as TeV-level electron positron colliders. M. Chen and his colleagues proposed a compact and efficient scheme to realize multistage LWFA based on curved plasma channels [Fig.~\ref{fig:schematic}~(a)], where the electron beam and laser pulse are simultaneously coupled into the next stage with high capture efficiency~\cite{Luo} 
Such scheme benefits from a shorter coupling distance and continuous guiding of the electrons in plasma while suppressing transverse beam dispersion. The key challenge of this scheme is to stably guiding relativistic laser from a curved plasma channel to a straight plasma channel.  

Recently, M. Chen and his colleagues have experimentally demonstrated that intense laser guidance and wakefield acceleration can be realized using a centimeter-scale curved plasma channel~\cite{Zhu}. 
When the channel curvature radius is gradually increased and the laser incidence offset is optimized, the transverse oscillation of the laser beam can be mitigated. GeV-level electron acceleration by the guided laser pulse in the curved channel was experimentally observed [Fig.~\ref{fig:schematic}~(b)]. These results pave a way towards the realization of seamless multistage LWFA. 

\vspace{1cm}
\section{Hybrid LWFA-driven PWFA as a test platform for staged plasma acceleration \\ 
\small{\textit{S. Sch{\"o}bel et al. (Hybrid Collaboration)}}\normalsize}

\noindent \textit{Helmholtz-Zentrum Dresden – Rossendorf, Germany}
\\
\\
Staging, sequencing plasma accelerator modules, is an important concept for the further development of plasma-based acceleration methods to reach the energy level required for high-energy physic applications. The in- and out-coupling of drive laser beams as well as matching of the electron beam into the plasma cavity of a subsequent energy booster stage is very challenging, as it requires precise spatial-temporal overlap to achieve quality-preserving acceleration. Besides the demand of extremely tight tolerance in pointing stability, the ability to precisely monitor the process is required to gain the level of control needed. 
\begin{figure}[htb!]
    \centering
    \includegraphics[width=\columnwidth]{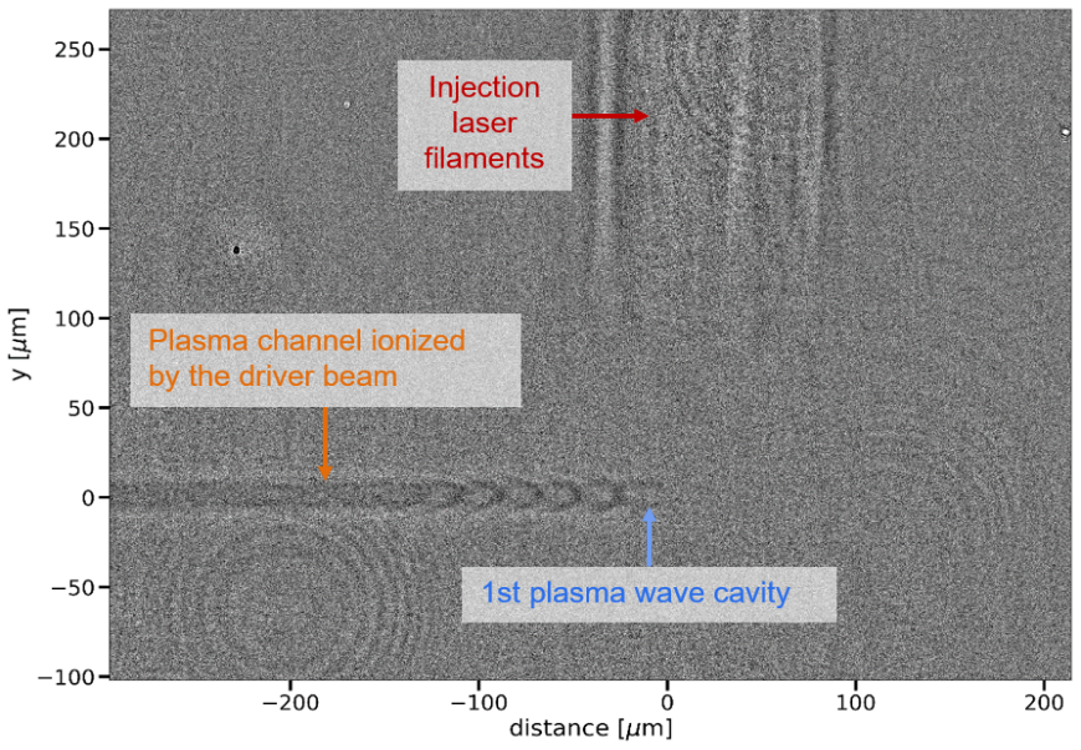}
    \caption{Shadowgraphy image obtained using few-cycle probing during the alignment process of Trojan Horse Injection in a PWFA stage driven by an LWFA generated beam}
    \label{fig:summary}
\end{figure}

In our recent hybrid LPWFA experiments, the challenge of a precise spatial-temporal overlap is addressed in the framework of the realization of Trojan Horse Injection in the PWFA stage. Using the multi-beam capability of the DRACO laser system, we demonstrate the overlap of beams generated from the two different arms of the system. These are split from a common source roughly 80m before the target, passing separate amplifiers and compressors. In the experiment, the 150\,TW arm of DRACO is used to generate high peak current electron beams in an LWFA stage, which are transported into a downstream plasma module to excite plasma waves. There, an injector laser, extracted from the other arm (1\,PW line), is used to release electrons via ionization directly inside the beam driven cavity, forming the witness beam which is to be accelerated in the PWFA stage. Thus, to consistently inject electrons, stable spatial (micrometer scale) and temporal (fs scale) overlap is required. To track and control the position and arrival time of both, the electron beam and the injector laser, few-cycle laser shadowgraphy turned out to be a key diagnostic. Since probing is an in situ, non-invasive technique, it can be used throughout the whole experiment, allowing not only precise initial alignment of all beams, but also the analysis of jitters and drifts of the single components. Since this technique is applicable in a wide density range ($10^{17}-10^{19}\,\mathrm{cm}^{-3}$), it can be relevant also for LWFA-LWFA staging.

Using this setup, we demonstrate Trojan Horse injection to be realized. Within 92\% of 50 consecutive shots we reached a successful injection of a witness beam, showing a sufficient overlap of the cavity and the injector laser. So far, the quality of the injected witness bunches jitters due both, alignment jitters of the cavity and the injector laser as well as fluctuations of the charge of the driver beam. Nonetheless, some beams show a very promising improvement in beam quality, for example a narrow energy spread (1.5\%). For these witness bunches, the charge density (charge per energy slice) is even exceeding the one of the driver beam.

This experiment therefore demonstrates the ability to control positioning of the injector laser aided by ultra-fast optical probing as well as the potential of Trojan Horse injection to serve as a quality booster. Our unique infrastructure is also applicable for multiple LWFA stages, thus providing a test-bed for staging.

\vspace{1cm}
\section{Simulations of Next-Generation Colliders\\ 
\small{\textit{A. Huebl, R. Lehe, R. T. Sandberg, M. Garten, A. Formenti, O. Shapoval, C. E. Mitchell, C.  Benedetti,  and  J.-L.  Vay}}\normalsize}

\noindent \textit{Lawrence Berkeley National Laboratory, Berkeley, USA}
\\
\\
As part of the ALEGRO (Advanced LinEar collider study GROup) Workshop 2024, we presented an invited talk on simulations of staged plasma accelerators towards future colliders. The talk built on theoretical foundations for staging \cite{sim_1,sim_2,sim_3,sim_4} and covered algorithmic options for modeling, the need for a compatible ecosystem of simulation codes and recent numerical modeling results.

\begin{enumerate}
\item Algorithmic Options were presented for modeling, from first principle simulations (with full, electro-magnetic particle-in-cell) to effective approximations and data (ML) models. There are general algorithmic choices to make between speedy simulations, which are fast and as accurate as possible, and high fidelity simulations, which are accurate and as fast as possible. The former include reduced physics (e.g., quasi-static and electro-static approximations, cylindrical geometry, fluid backgrounds) and the latter high-resolution, 3D3V, electromagnetic modeling, with a near-continuum of hybrid models in between. Reduced physics models are required for rapid initial designs, optimizations and operations. Full fidelity physics models are needed for stability proofs, exploration and ML training data generation.

From particle source over staged acceleration to interaction point the modeling requirements vary: plasma source/beam generation often requires full electromagnetic PIC \cite{sim_5, sim_6}, potentially with moderate boosted frame (e.g., $\gamma = 5$). Staged plasma acceleration benefits from full PIC with high  relativistic $\gamma$ factor or quasi-static codes \cite{sim_7}. Transport is best performed with electrostatic PIC in the beam frame with s-based modeling \cite{sim_8, sim_9}. Interaction point physics at ultra-relativistic energies can use electrostatic PIC to model beam crossing coupled with Monte-Carlo QED modules \cite{sim_10}. A compatible ecosystem of codes, implementing and sharing models and data, needs to be striven for that uses standardized input/output and common principles/practices (e.g., open source development practices, continuous integration testing/benchmarking, open documentation) \cite{sim_11, sim_12}.

\item	Community Ecosystems. The Beam, Plasma \& Accelerator Simulation Toolkit (BLAST) was presented as a compatible toolkit striving to address these modeling needs from laptop to Exascale supercomputer \cite{sim_6,sim_7,sim_8,sim_9, sim_13,sim_14,sim_15,sim_16}. BLAST codes are part of the Collaboration for Advanced Modeling of Particle Accelerators (CAMPA), which is highly synergistic with ALEGRO goals to design advanced plasma-based colliders. CAMPA also develops novel algorithms, supports standards such as the Particle-In-Cell Modeling Interface (PICMI) and open particle-mesh data standard (openPMD) \cite{sim_17,sim_18,sim_19}, codes beyond BLAST \cite{sim_20, sim_21}, laser manipulation and exchange (LASY) \cite{sim_22, sim_23}, and ML-based optimization (optimas) \cite{sim_24, sim_25}.

\item	Modeling Staging: Levels of Realism. One of the pressing needs of the community towards trustworthy plasma-based collider designs is to systematically increase the realism via start-to-end modeling \cite{sim_26}. That requires stepwise maximizing energy gain while conserving transported charge, minimizing energy spread, controlling emittance growth, and ultimately ensuring compactness and energy efficiency as well as robustness under realistic profiles \cite{sim_22}, fluctuations and uncertainties in operations. In modeling, this requires establishing work-flows (e.g., optimization \cite{sim_24}) that are easy to reproduce, automate \& repeat, memorize (with ML) \cite{sim_27, sim_28}, and abstract away.

For acceleration stages, 3D WarpX simulations with low witness beam charge were presented, increasing the currently modeled number of stages from 3 to 50 \cite{sim_29, sim_30}. Individual stages were then optimized with electrostatic RZ modeling \cite{sim_13} using ML-guided optimization (Bayesian Optimization) \cite{sim_24, sim_31} to find LPA downramp profiles below the adiabatic limit for stages from 1 GeV to 10 TeV while preserving emittance growth up to 10 pC. Addressing a need to model the plasma-conventional hybrid beamlines required for transport gaps in a collider, a novel surrogate approach for including plasma elements in beamline modeling was explored \cite{sim_28}. In the presented approach, high-fidelity, full PIC simulations (WarpX) were used to train a neutral network \cite{sim_27} that then enabled \%-level accurate tracking of beam moments using ML inference of trained LPA stages via all-GPU accelerated ImpactX beamline simulations. The achieved performance for GPU inference was 63\,ns / particles / stage with total simulation runtime of 15 stages and transport as low as 2-4 simulations per GPU and second. It is envisioned that this will enable rapid design studies of complex transport gaps, e.g., for HALHF \cite{sim_32, sim_33}. Lastly, 3D WarpX simulations and new collaborations were presented to study beam crossing for machines such as ILC, HALHF and others \cite{sim_33,sim_34,sim_35}.

\end{enumerate}
  
\vspace{1cm}
\section{A Hybrid, Asymmetric, Linear Higgs Factory (HALHF)\\
\small\textit{R. D'Arcy\textsuperscript{1}, B. Foster\textsuperscript{1,2}, and C.A. Lindstr{\"o}m\textsuperscript{3}}\normalsize}

\textit{\textsuperscript{1}Oxford University, Oxford, United Kingdom}\\
\textit{\textsuperscript{2}DESY, Hamburg, Germany}\\
\textit{\textsuperscript{3}University of Oslo, Oslo, Norway}
\\
\\
Construction of a Higgs factory is the top priority for particle physics in the next decades but the costs are extremely high. Plasma-wakefield accelerators (PWFAs) promise to reduce drastically the footprint and cost of such machines. However, while progress on electron acceleration is rapid, positron acceleration in plasma remains challenging due to the inherent charge asymmetry in plasma. We propose a pragmatic approach to a linear-collider concept that bypasses the positron problem by using a hybrid of classical radio-frequency (RF) accelerators to accelerate positrons and PWFAs to accelerate electrons aka HALHF~\cite{DArcy_1}.

This hybrid scheme requires the beam energies to be highly asymmetric in order to prevent the facility cost being dominated by the classical positron linac. A sensible choice of beam energies is motivated by a reduction by 4x in positron energy (31 GeV) and an increase by 4x in electron energy (500 GeV) compared to symmetric beams producing 250 GeV centre of mass. GUINEA-PIG simulations give a similar geometric luminosity to that of the International Linear Collider (ILC) 250 GeV scheme but HALHF requires $\approx$2x more power to produce it. Asymmetric bunch charges can compensate for this: 4x more positrons and 4x fewer electrons per interaction ideally but, given the difficulty of producing positrons, a factor of 2 was used instead. Finally, as the geometric emittance of the beams scales inversely with energy, it is possible to ease the emittance requirements of the plasma-accelerated electron beam by decreasing the $\beta$ function at the interaction point. In short, HALHF benefits from asymmetry in many key areas.

By utilising a 400 m plasma linac to accelerate electrons to 500 GeV with an accelerating gradient of 6.4 GV/m the facility footprint is 3.3 km in length, substantially reduced from that of e.g. the ILC. As such, HALHF would fit on most major pre-existing particle physics laboratories. Furthermore, the build cost of HALHF is estimated at a total cost of \$1.9B, greatly reduced compared to the ILC.

Since publication of the original paper~\cite{DArcy_1}, the authors of the HALHF concept have proposed a  suite of staged facility upgrades and costings~\cite{DArcy_2} to the original HALHF concept. These include: a polarised positron source via an ILC-like scheme but benefitting from minimally disrupted electron beams; an centre of mass boost to 380 GeV to access the t-tbar threshold, as with the CLIC baseline design; two interaction points to minimise systematics with two complementary detectors and to therefore double the luminosity; and an upgrade to 1 TeV centre of mass by colliding $\gamma$ beams produced via Compton scattering (either with an optical laser or integrated XFEL) with two 500 GeV electron beams.

The wider community has come together to assess the outstanding challenges preventing the realisation of a self-consistent Conceptual Design Report (CDR). This has taken the form of a series of in-person and online meetings, beginning with an in-person ‘Kick-off’ meeting at DESY in Oct 2023, proceeding with monthly online meetings, and most recently with an in-person workshop in Oslo. The major challenges identified on the plasma side—beyond the obvious need to stage plasma sources to reach very high energy and operate those plasma sources at MHz repetition rates \& 10 kW/m average powers to achieve the required luminosity—are large transverse instabilities (leading to large emittance growth) and beam-induced ionisation of higher levels of argon (significantly modifying the on-axis plasma density between acceleration events). These can be mitigated by a reduction in operational plasma density (and therefore beam density due to the elongation of the beam)—a decision that was recently made at the Oslo Workshop, reducing the plasma density by approximately an order of magnitude to $10^{15}~\mathrm{cm}^{-3}$. Another challenge to operation of the plasma linac was recently discovered by HALHF collaborators, namely the difficulty of preserving the flat transverse beam profile (required to maximise luminosity and minimise beamsstrahlung at the interaction point) in a plasma due to resonant emittance mixing between the two transverse planes in the presence of e.g. ion-motion-induced field non-uniformities~~\cite{DArcy_3}. Beyond the plasma challenges, the question of which classical RF technology to use for the combined positron/electron-drive-beam linac remains open, with neither option of a CW SCRF or warm L-band linac appearing easy, and the positron source remains an open challenge.

The next steps for the HALHF project are to work towards a self-consistent parameter set required to generate a ‘pre-CDR’ for input to the European Strategy for Particle Physics (ESPP) Roadmap Process, as well as to the comprehensive global Linear Collider plan, in March of 2025. The monthly meetings will therefore continue in earnest with an ‘Experts Meeting’ to make progress in finalising the ‘HALHF 2.0’ parameters, expected to take place in Erice, Sicily in October of this year.
\vspace{1cm}
\section{Preliminary Investigation of a Higgs Factory based on Proton-Driven Plasma Wakefield Acceleration \\
\small{\textit{J. Farmer\textsuperscript{1}, A. Caldwell\textsuperscript{1}, and A. Pukhov\textsuperscript{2}}}\normalsize}

\textit{\textsuperscript{1}Max-Planck-Institut für Physik, München, Germany}\\
\textit{\textsuperscript{2}Heinrich-Heine-Universität Düsseldorf, D{\"u}sseldorf, Germany}
\\
\\
With the discovery of the Higgs Boson in 2012, attention in the high-energy physics community has turned to particle colliders that would allow its detailed study. Here, we discuss the possibility of a proton-bunch driven plasma-wakefield acceleration scheme \cite{apukhov_1}. We find that such a scheme is an attractive option as 1) it is more compact than the conventional RF acceleration schemes, and 2) is simpler than the electron-driven PWFA scheme as staging is not necessary and as both positrons and electrons can be accelerated in the proton-driven wakefield, allowing for a symmetric collider. Significant challenges are present, however, in providing the desired luminosity. As such, our work provides a conceptual basis for such a facility in the hope that the challenges described can be met. 

\begin{figure}[htb!]
    \centering
    \includegraphics[width=0.5\columnwidth]{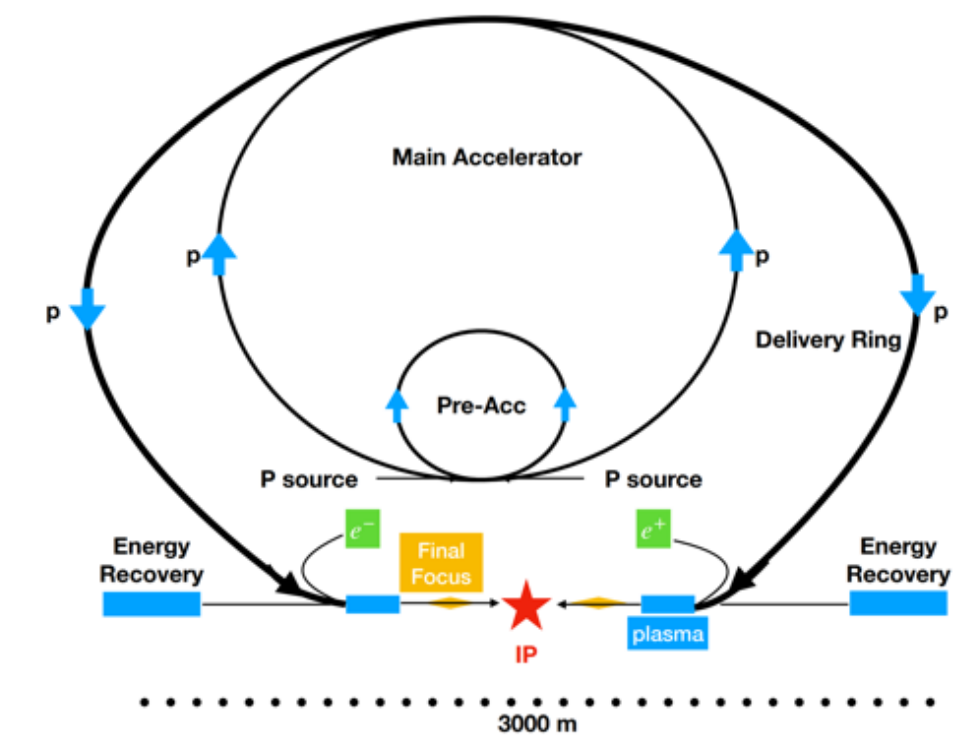}
    \caption{Schematic accelerator layout for our concept. A fast-cycling proton synchrotron, fed by a series of pre-accelerators, provides the drive for the plasma wakefields. The radius of the largest proton accelerator is approximately 1000\,m, while the plasma acceleration stages are about 150\,m-long. A final focusing region is indicated, as well as energy recovery plasma sections.}
    \label{fig:schematicP}
\end{figure}

The advantage of a proton driver is that the energy of the driver, using today’s technology, is sufficient to easily reach accelerated bunch particle energies of the required level for a Higgs Factory without staging, greatly simplifying the accelerator complex. A limitation of the proton-driven scheme has been the repetition rate of the driver, which limits the achievable luminosity. With the development of fast-ramping superconducting magnets \cite{apukhov_2, apukhov_3}, this limitation can be largely removed, making the proton-driven scheme very attractive. A schematic layout for the prospected accelerator is shown in Fig.~\ref{fig:schematicP}.

We have simulated electron and positron acceleration in our proton-driven scheme, assuming that a suitable driver would be available. The simulations are preliminary as the required numerical resolution is not available in today’s codes. However, the simulations do provide proof-of-principle for the energy gain, and give a basis for the expectations of what could eventually be achievable. We then evaluate the luminosity for a Higgs Factory based on our set of assumptions for the proton-driven acceleration complex and the plasma-based acceleration stage. Final focus parameters are taken from other studies. We conclude with a discussion of the developments that would be necessary to make this approach reality.

\vspace{1cm}
\section{Resonant emittance mixing of flat beams in plasma accelerators \\
\small{
\textit{
S.~Diederichs\textsuperscript{1,2}, C.~Benedetti\textsuperscript{3}, A.~F.~Pousa\textsuperscript{1}, A.~Sinn\textsuperscript{1}, J.~Osterhoff\textsuperscript{1,3},\\ C.~B.~Schroeder\textsuperscript{3,4}, and M.~Thévenet\textsuperscript{1}
}}\normalsize}

\textsuperscript{1}Deutsches Elektronen-Synchrotron DESY, 22607 Hamburg, Germany\\
\textsuperscript{2}CERN, Espl. des Particules 1, 1211 Geneva, Switzerland \\
\textsuperscript{3}Lawrence Berkeley National Laboratory, 1 Cyclotron Rd, Berkeley, California 94720, USA\\
\textsuperscript{4}Department of Nuclear Engineering, University of California, Berkeley, California 94720, USA
\\
\\
Plasma-based accelerators are promising candidates as drivers for future linear colliders due to their $\gtrsim$ GV/m accelerating gradients. 
In a linear collider, the luminosity $\mathcal{L}$ must be maximized while simultaneously minimizing deleterious beamstrahlung effects.
A common solution is to operate with flat beams, $\sigma_x \gg \sigma_y$ and thus $\epsilon_x \gg \epsilon_y$, where $\epsilon_{[x,y]}$ is the beam emittance in $[x,y]$. This ratio should be preserved during acceleration, i.e., emittance mixing should be avoided.
In general, emittance mixing occurs when the equations of motion in $x$ and $y$ are coupled (for instance when the transverse force in $x$ depends on $y$). Here, we briefly summarize the findings on emittance mixing in plasma-based accelerators in \cite{Diederichs:2024}.

Plasma accelerators are often operated in the so-called blowout regime, where the driver is strong enough to expel all plasma electrons, creating a trailing ion cavity in its wake. For a uniform background ion distribution within the cavity, the transverse wakefields in $x$ and $y$ are decoupled, preventing emittance exchange. In practice, various nonlinear effects can perturb the transverse wakefields and cause coupling and, hence, emittance mixing. Such effects are almost inevitable for collider-relevant beams that require high charge ($\sim$nC) and low emittance ($\sim100$ nm) and therefore generate extreme space-charge fields. These fields can further ionize the background plasma to higher levels and/or cause ion motion, both of which can lead to the formation of nonlinearly coupled wakefields. 

\begin{figure}[htb!]
    \centering
    \includegraphics[width=1.0\columnwidth]{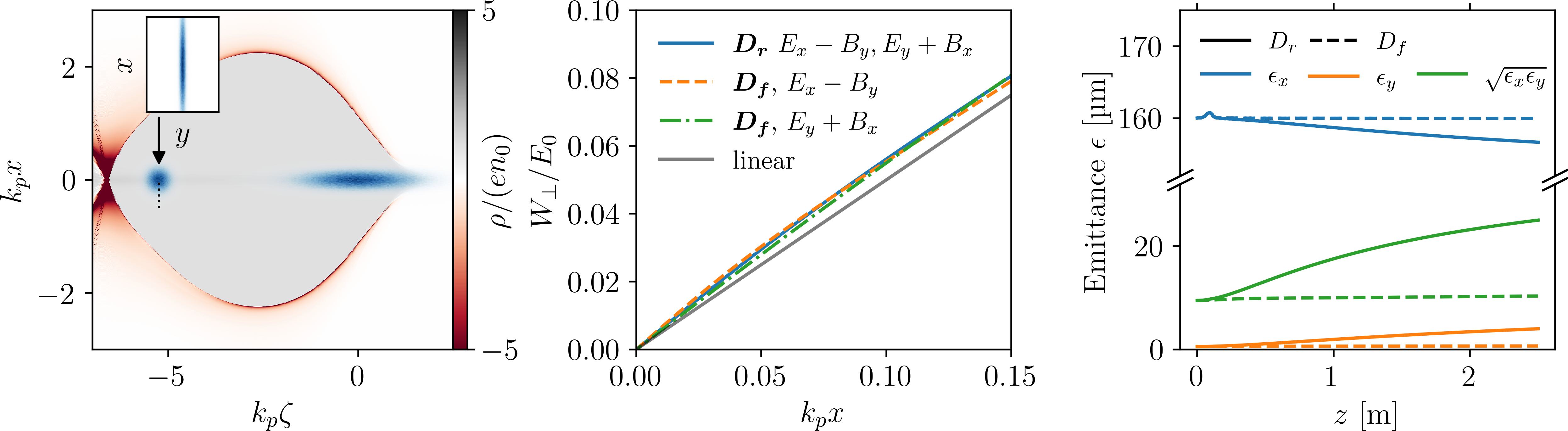}
    \caption{Left: Blowout wake with mild drive beam ion motion. Middle: Transverse wakefields for a round (blue) and a flat (orange, green) driver. Right: Emittance evolution for a flat witness beam with a round driver (solid lines) and a flat driver (dashed lines), and for a round witness with a round driver.}
    \label{fig:Fig:1}
\end{figure}

The coupled, nonlinear wakefields in plasma accelerators can lead to severe emittance mixing of flat beams when there is a resonance between the betatron oscillations of beam particles in the horizontal and vertical planes, as shown in Fig.~\ref{fig:Fig:1}. With this effect, the emittance in $x$ decreases as the one in $y$ increases, resulting in an overall growth of their geometric average $\sqrt{\epsilon_x \epsilon_y}$. 

While a radially symmetric nonlinear wakefield (here caused ion motion induced by a round driver $D_r$) always results in full emittance exchange, breaking the radial symmetry (e.g., by using a flat driver $D_f$) mitigates this effect by detuning the resonance.  
However, even for non-radially symmetric wakefields emittance mixing can have a significant effect. Furthermore, emittance mixing can also occur in other regimes such as the linear and quasi-linear regimes as well as plasma-based positron acceleration schemes that operate with coupled, nonlinear focusing fields.
Consequently, emittance mixing is an important effect and must be taken into account when designing plasma-based colliders using flat beams.

\vspace{1cm}
\section{Advancements in Beam Delivery Systems: CLIC Innovations and Plasma Collider Applications \\
\small{\textit{V. Cilento}}\normalsize}

\textit{CERN, Geneva, Switzerland}
\\
\\
\subsubsection*{Introduction and Objectives}
The presentation by Vera Cilento, Enrico Manosperti, and Rogelio Tomás explores advancements in the beam delivery systems (BDS) focusing on the Compact Linear Collider (CLIC) and plasma collider applications. It elaborates on CLIC’s BDS components such as the Diagnostic Section, Collimation Section, and Final Focus System, which are essential for minimizing beam size, correcting chromatic aberrations, and maintaining beam stability at energies of 380\,GeV and 3\,TeV.

\subsubsection*{Dual BDS Concept}
The dual BDS concept for CLIC, designed for up to 3\,TeV, addresses challenges posed by larger crossing angles and higher energies. It aims to balance a total luminosity loss of about 30\% for extra lines with chromaticity correction and minimal trajectory bending to reduce synchrotron radiation effects.

\subsubsection*{Scaling with Energies}
Future BDS designs aim to scale to higher energies, as shown in the proposed 7\,TeV CLIC BDS design. This design focuses on improving transverse aberration control at the interaction point (IP) and aims for a target luminosity of approximately $10^{35} \, \text{cm}^{-2} \text{s}^{-1}$, taking into account variations due to the solenoid field.

\subsubsection*{Synergies with Plasma Colliders}
Collaborative efforts with plasma collider projects such as Laser Plasma Accelerators (LPA) and Hybrid, Asymmetric, Linear Higgs Factory (HALHF) have identified shared solutions for common challenges. These collaborations are critical for achieving desired luminosity goals while addressing issues like emittance preservation and energy spread.

\subsubsection*{Conclusions}
The presentation concludes with the importance of BDS innovations in future collider designs. It emphasizes that exploring scaling laws and collaborations with plasma colliders are crucial for effectively tackling the demands of higher energy levels, enhancing overall collider performance.

\vspace{1cm}
\section{Laser-driven production of ultra-short high quality positron beams\\
\small{\textit{G. Sarri}}\normalsize}

\textit{Queen's University, Belfast, UK}
\\
\\
Wakefield acceleration of electrons has achieved a relatively high level of maturity, with several landmark results reported both in laser-driven and beam-driven configurations. However, wakefield acceleration of positrons is still at its infancy, a major issue for the design of the next generation of lepton colliders based on plasma technology. The considerably higher complexity of accelerating positrons in a plasma cavity can be easily understood if one considers the intrinsic structure of wakefield accelerators, which, for positively charge particles, tend to be defocusing and to have extremely narrow accelerating regions.

Several alternative schemes have been numerically and theoretically proposed (including hollow plasma channels~\cite{sarri_1} and finite plasma columns~\cite{sarri_2}) and a few proof-of-principle experiments have shown potential in this direction (e.g.,~\cite{sarri_3,sarri_4}). However, progress in this area is currently significantly hampered by the lack of experimental facilities able to provide positron beams suitable to be guided and accelerated in a wakefield accelerator, with only FACET-II at SLAC suited in principle for this type of work~\cite{sarri_5}. To address this need – also identified in several international roadmaps and strategy documents (e.g.,~\cite{sarri_6,sarri_7}) – positron wakefield acceleration is now included in the scientific mission of several plasma-accelerator facilities (e.g,~\cite{sarri_8,sarri_9}). 

We recently demonstrated experimentally~\cite{sarri_10} that high-intensity lasers of relatively modest peak power (100\,TW) can indeed produce positron beams with appealing characteristics, including tuneable energy (100 - 600\,MeV), low geometrical emittance ($\simeq$15\,nm), and short duration ($\simeq$~10\,fs). Moreover, the first implementation of a simple magnetic beamline has demonstrated energy selection of beamlets with <5\% energy spread~\cite{sarri_10}. Proof-of-principle Particle-In-Cell simulations have shown that beams of this kind are already of sufficient quality to be trapped and accelerated in a wakefield accelerator~

Due to the relatively low power of the laser beam, the overall charge achieved did not allow for beam-loading in the cavity. However, numerical scaling of the experimental results to PW-class lasers has shown that GeV positron beams with micron-scale normalised emittance and a charge in the region of a few pC in a 5\% bandwidth can be achieved~\cite{sarri_11}. This has been numerically demonstrated for the EuPRAXIA laser facility, which already includes a design for a beamline capable to energy-select and focus the positron beams down to 20\,micron~\cite{sarri_8}. Such positron beams would be ideal to act as witness beams in a facility suitable for positron wakefield acceleration studies and, therefore, provide a unique capability for the development of plasma modules to be included in the design of the next generation of particle colliders. Further scaling to the 10 PW level is also numerically showing that a further increase in charge up to 30-50\,pC charge in a 5\% bandwidth can be obtained, while preserving femtosecond-scale duration and $\mu$m-scale normalised emittance at the GeV level.
\vspace{1cm}
\section{Experience with Wakefield Acceleration at SwissFEL\\
\small{\textit{E. Ericson}}\normalsize}

\textit{Paul Scherrer Institute, École Polytechnique Fédérale de Lausanne, Wurenlingen, Switzerland}
\\
\\
SwissFEL is a hard X-ray Free electron laser with hard and soft X-ray beamlines. The accelerator is over 700/,m long and produces beams above 6/,GeV with remarkable stability. The stability is quantified in terms of energy jitter, arrival time jitter and trajectory. The EuPRAXIA project would like to reproduce this beam quality using a plasma accelerator however plasma accelerators currently struggle to reliably produce beams with small energy spread due to the short wavelength of the plasma wake. One proposed solution to this problem is to use passive wakefield structures to manipulate the energy spread of the beam. Passive structures are classified by their geometry and material. 
\newglossaryentry{PSI}{name={PSI},description={Paul Scherrer Institute}}
PSI has flat dielectric structures, rectangular metallic structures, and round dielectric structures. Round corrugated structures are studied at other laboratories. PSI currently has two flat dielectric passive structures installed with alternating orientations after LINAC3. They are routinely used to change the bandwidth of the FEL and create ultrashort X-ray pulses.  

My work has been to analyze the results of previous experiments using the flat dielectric passive structures and explain the observations using simulations. The first experiment I considered is a single-bunch experiment where the flat structure is used as a dechirper. A beam with an energy chirp such that the tail has a higher energy than the bunch head is sent into the structure and the wakefields act back on the beam to reduce its energy spread while also changing the central energy of the beam. This effect is amplified when the structure has a small gap.  

In the experiment, the opposite was done: a monochromatic beam was sent into the structure and the chirp of the beam exiting the structure was measured. The energy spread of beams with three different lengths were measured as the structure gap was varied. The evolution of the beams’ energy spreads and centroid energies for different gap settings is apparent when looking at the beam images.  

To explain the observed energy spectra of the beams exiting the flat dielectric passive structure, I simulated a model of the flat dielectric passive structure in CST Studio. I calculated the wake potentials of relatively long beams (0.3\,mm < $\sigma<$ 1.1\,mm) for ten different gap settings between 0.5\,mm and 3.0\,mm. I calculated the loss factors, a quantity that predicts the change in energy of the beam due to the wakes, for each bunch length and structure gap setting. To calculate the loss factors of the bunches with $\sigma\approx$mm used in the experiment, I used ECHO. After benchmarking ECHO against CST and performing convergence studies of the mesh and simulation settings, I computed the wake potential for the three different bunch lengths as a function of the structure gap. While the error analysis for the experiment still needs to be done, I have reproduced the loss factors measured during the experiment using wake simulations.  

A second experiment was done using the flat dielectric passive structure: two bunches were sent into the structure with a small delay between them and the beam energies were measured. When the structure gap and delay between bunches are large, the beam image showed a single group of charge. When the experiment parameters were varied, we observed two distinct charge groupings: one at the initial energy and a second, less intense bunch, with a higher energy. The measurements suggest we have used a 200\,pC drive beam to accelerate a 25\,pC witness beam by transferring energy from the drive beam to the witness beam. The measurements show this process is sensitive to the spacing between the two bunches and the structure gap. The accelerating gradient in the structure and cause of the large energy spread observed for small structure gaps are still unknown. 

Going forward, I will calculate the energy spread as a function of the structure gap for the single-bunch experiment and analyze the measurements done during the two-bunch experiment and perform long-range wakefield simulations to understand what accelerating gradients are possible in the structure. I will use the SwissFEL beam to understand the performance of wakefield structures to improve the quality of beams produced by plasma accelerators. 
\vspace{1cm}
\section{Six-dimensional phase space preservation in a terahertz-driven multistage dielectric-lined rectangular waveguide accelerator\\ 
\small{\textit{O. Apsimon}}\normalsize}

\textit{University of Manchester, Manchester, U.K.}
\\
\\
The realization of laser-driven high-field THz sources \cite{oapsimon_1} led to several concepts being proposed for the use of THz radiation for the acceleration and manipulation of electron beams, including interactions with two counterpropagating THz pulses \cite{oapsimon_2,oapsimon_3}, in circular dielectric-lined waveguides \cite{oapsimon_4,oapsimon_5}, using a subluminal travelling terahertz source \cite{oapsimon_6} and high-power narrow-band THz generation \cite{oapsimon_7}. A concept has also been proposed for using an evanescent-wave scheme to accelerate protons \cite{oapsimon_8}. THz acceleration has been experimentally demonstrated at nonrelativistic \cite{oapsimon_9,oapsimon_10,oapsimon_11,oapsimon_12} and, more recently, relativistic beam energies, using either a laser-driven THz source \cite{oapsimon_13} or a coherent-transition-radiation-based THz source \cite{oapsimon_14}. Frontier studies focus on the stability and scalability using multistaging \cite{oapsimon_15}. The first demonstration of laser-driven THz acceleration of relativistic electrons was conducted in 2019 using the Compact Linear Accelerator for Research and Applications (CLARA) test facility at Daresbury Laboratory \cite{oapsimon_13}. This experiment used a dielectric-lined rectangular waveguide structure driven by narrow-band, frequency-tunable, polarization-tailored THz pulses and demonstrated up to 10 keV acceleration of a 2 ps (subcycle) bunch with an initial energy of 35 MeV. Consequently, it marked a key milestone on the path to whole-bunch linear acceleration of subpicosecond electron beams with multistaged concepts capable of preserving beam quality.

The control and preservation of beam quality are central to the success of future THz-driven accelerators. However, solutions to implement them in a scheme that preserves beam properties have yet to be studied in detail. The figure of merit for beam quality is six-dimensional (6D) phase-space brightness which includes both transverse emittance and energy spread. For a beam with a finite longitudinal size, both quantities can be defined as a thin longitudinal slice or the entire bunch of a beam as a projection of all slices. In some cases, where nonlinear forces are exerted on the beam, such as space charge defocusing or the time variation of an accelerating field across a bunch, the projected quantities can be large even though each slice has a high 6D brightness. Consequently, it is relevant to distinguish both definitions and characterize the 6D phase space for each, especially for applications such as free electron lasers where slice properties are more relevant due to the microbunching process dominating the radiation emission characteristics. Previously, synchronous acceleration of nonrelativistic electrons in a tapered dielectric-lined waveguide structure \cite{oapsimon_16}, the effect of the longitudinal focus position and phase stability on the beam dynamics \cite{oapsimon_17} and compensation of nonlinear energy chirp due to radio-frequency field curvature in corrugated structures \cite{oapsimon_18} have been studied.

In this contribution, we demonstrated that a dielectric lined rectangular waveguide can preserve both emittance and energy spread, simultaneously, using an orthogonal multistaging scheme with the implementation of appropriate multistaging coupling. We have demonstrated the concept of correlated energy spread correction via an orthogonal multistaging of dielectric-lined rectangular waveguide structures. To do this, the longitudinal multipole component of a THz field induced in a single structure was decomposed and the resulting transverse voltages were studied. It was shown that, for a THz field synchronous with electrons, a rectangular DLW, unlike conventional RF structures, lacks a monopole component of the transverse voltage, which is commonly a large contribution to the emittance growth of the beam during acceleration. Therefore, we have achieved the preservation of slice properties in the six-dimensional phase space both correcting for the induced correlated energy spread up to 82\% while preserving the injected transverse emittance. To study the interaction of a relativistic beam with the decomposed multipole components of the THz field and associated transverse voltages, a multipole-based tracking code was developed and benchmarked against CST Studio Suites’s PIC solver. This improved the computing speed significantly allowing global characterization of beam dynamics under THz fields. Finally, we have studied the effect of a finite bandwidth THz field similar to that expected from real THz sources by demonstrating that the transverse voltage due to two sidebands around the central frequency mostly cancels, leaving a residual emittance growth of only 0.2\% (within the statistical error of 0.3\%). The energy spread correlated to the THz field also remains identical to the single frequency case. 

This is a summary from \cite{oapsimon_19} more details can be found in this reference and references therein. 
\vspace{1cm}
\section{AWAKE: a plasma wakefield accelerator for particle physics\\ 
\small{\textit{M. Turner et al., AWAKE Collaboration}}\normalsize}

\textit{CERN, Geneva, Switzerland}
\\
\\
To effectively accelerate electrons to energies relevant for particle physics applications in plasma wakefields, drivers require: 
\begin{itemize}
    \item beam densities similar to or exceeding the plasma density or laser peak intensities with $a_0\sim1$ or larger,
    \item lengths on the order of the plasma wavelength (\textmu m-mm),
    \item a high energy per driver (kJ) and in the case of beam drivers also per particle (few 10s-100s\,GeV), and 
    \item a high group velocity in plasma ($v_g \sim c$).
\end{itemize} 
Currently, no drivers with all these properties exist. %
Many experiments use either short and dense particle drivers which are limited in energy or short and intense laser drivers, which have a low group velocity in plasma. %
Therefore, tens to hundreds of plasma stages with separate drivers are required to reach energies relevant for particle physics, posing significant challenges. %

Another possibility is developed as part of the AWAKE experiment at CERN, which is to use proton bunches as drivers. %
Available proton drivers have sufficiently high energy and phase velocity to enable particle physics relevant electron energy gains in a single plasma. %
However, they are of relatively low density and have lengths longer than the plasma wavelength. %
Using them at densities suitable for high gradient acceleration requires a more complicated excitation scheme, where the bunch is transformed into a train of micro-bunches and wakefields are excited resonantly using multiple micro-bunches. %
Using this scheme, AWAKE has a concrete plan to provide 50-200\,GeV electron beams with sufficient quality for first particle physics experiments in the early 2030’s. %

\begin{figure}[htb!]
    \centering
    \includegraphics[width=\columnwidth]{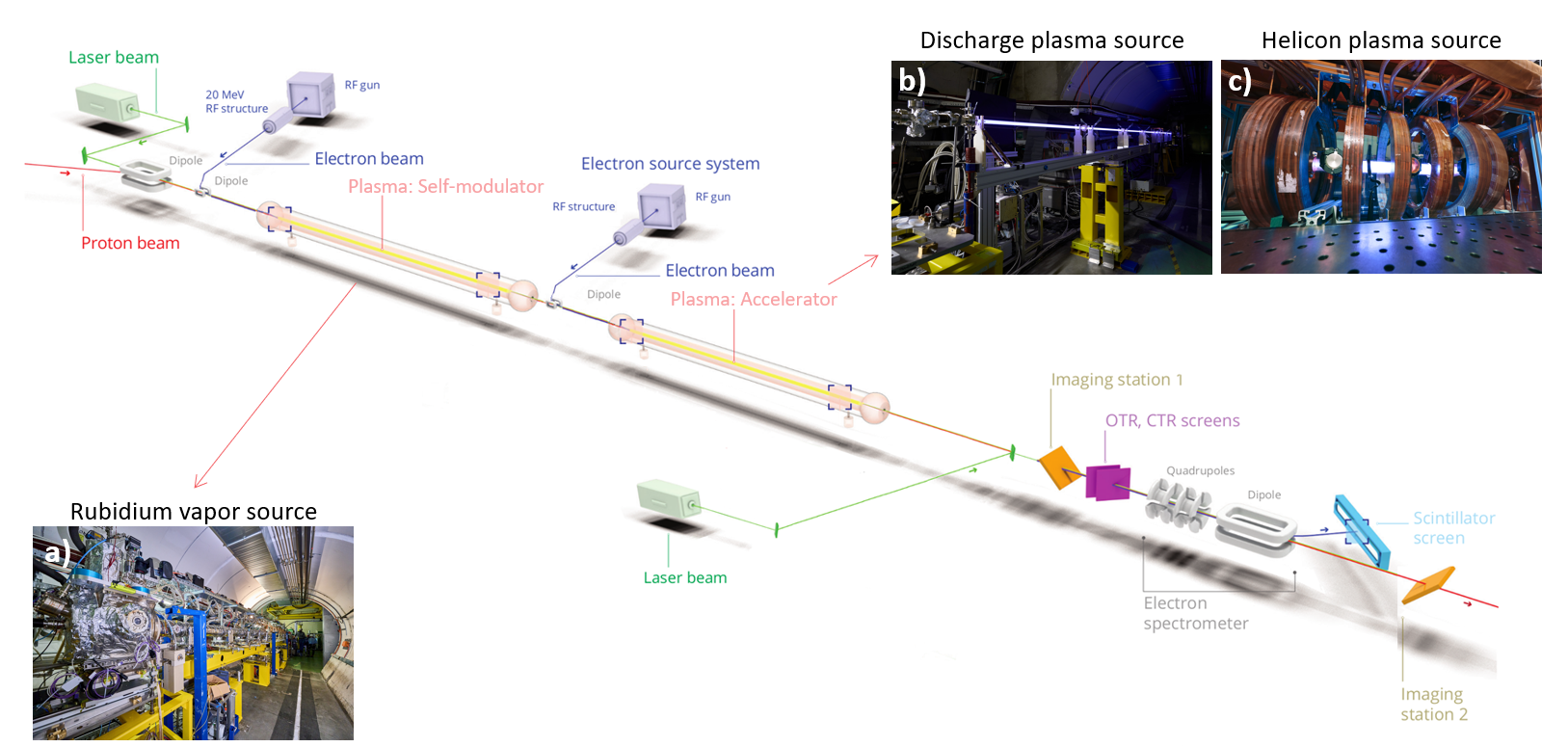}
    \caption{Schematic experimental layout for the AWAKE Run 2d program. The first plasma is the self-modulator; here the microbunch train is formed. In the current baseline, this plasma is a rubidium vapor source with a density step (a), but may also be of different technology. The second plasma is the accelerator; here witness particles are accelerated until their desired energy is reached. To be extendable, this plasma should be made by scalable technology (e.g. discharge plasma sources (b) or helicon plasma sources (c)). Bunches and pulses propagate to the right.}
    \label{fig:schematicAWAKE}
\end{figure}

AWAKE~\cite{bib:AWAKE} is located at CERN and uses single proton drive bunches from Super Proton Synchrotron (SPS). %
Proof-of-principle demonstration results have been completed between 2016 and 2018~\cite{NatureAWAKE2018,PhysRevLett.122.054802,PhysRevLett.122.054801,PhysRevLett.125.264801,PhysRevLett.126.164802}. %
Since 2021, AWAKE is progressing on its Run 2 program~\cite{sym14081680}, with the goal to transition the concept from proof-of-principle demonstrations to parameters ready for first particle physics applications, and to answer all questions connected to using the resonant excitation scheme. %
AWAKE Run 2 is split into four sub-phases, Run 2a to 2d. %
Run 2a was completed in 2022 with the demonstration that an electron bunch can be used as a seed for the self-modulation process~\cite{PhysRevLett.129.024802}. %
Run 2b (2023-2024) is ongoing and the goal is to demonstrate that a plasma density step allows to maintain and stabilize a high amplitude for the wakefields over long distances. %
For Run 2c (starting mid 2028), the goal is the demonstration of acceleration of a bunch with quality sufficient for first applications: 100\,pC, 2-30\,mm-mrad, and 5-8\% relative energy spread. %
This will be achieved by separation of self-modulation and acceleration. %
In preparation for Run 2c, AWAKE is making use of CERN’s long shutdown 3 to increase the size of the experimental facility, which is required to host all of Run 2c and d infrastructures. %
After demonstration of quality acceleration in Run 2c, Run 2d (schematic experimental layout on Fig.~\ref{fig:schematicAWAKE}) aims to demonstrate the scalability of the electron energy gain, while maintaining quality, by increasing the plasma length. %
This requires a scalable technology for the plasma source, for which AWAKE launched a dedicated plasma source R\&D program in 2018 with five collaborating institutes and CERN as a host. %
The program has two dedicated laboratories at CERN for development of helicon~\cite{Buttenschön_2018,{zepp}} and discharge~\cite{{Torrado}} plasma sources.

After completion of Run 2, expected in the early 2030s, AWAKE would be able to deliver $\sim$50-200\,GeV electron bunches for a first particle physics experiment. %
Options for applications include fixed target experiments to search for the dark photon or study nonlinear QED with electron photon collisions. %
On the longer timescale (and after further improvements in accelerated bunch quality), the use of 7\,TeV LHC drivers to reach TeV electron energies for $ep$ or $eA$ collisions is under consideration. %

AWAKE aims to serve particle physics applications and is therefore well integrated into the ESPP process. %
Additionally, AWAKE is well integrated into the plasma wakefield community since the experiment develops physics understanding and technologies relevant all plasma-based particle accelerators, such as e.g. external injection or plasma source development. %

\vspace{1cm}
\vspace{1cm}
\section{Positrons at FACET-II: Status and Potential\\
\small{\textit{M. J. Hogan}}\normalsize}

\textit{SLAC National Accelerator Laboratory, Menlo Park, CA, USA}
\\
\\ 
Linear Colliders collide electron and positron beams. They are used for precision particle physics studies. To date, there has only been one linear collider ever built: The SLAC Linear Collider (SLC) which operated from 1986-1998 with 100\,GeV pCM in 3km footprint. SLC technology does not scale well to today’s energy frontier (10\,TeV CM). 

Linear Colliders and their role in The Future of Particle Physics have been discussed in recent planning exercises in the U.S. and Europe. The U.S. Snowmass and European Strategy visions seem aligned – maximize HL-LHC and plan for a Higgs factory as a next step. The current vision of U.S. P5 (Particle Physics Project Prioritization Panel) aligns with Michael Peskin’s original ALEGRO provocation: we know how to build a Higgs factory with relatively mature technology (FCCee or ILC) and the AAC Community should focus on the energy frontier: 10\,TeV. The European situation is less binary though with the recent emergence and momentum of HALHF.
Collider concepts assume high degree of symmetry between electrons (e$^-$) and positrons (e$^+$), yet this is not a good assumption with plasma. To date, the only plasma acceleration experiments with e$^+$ have been conducted at SLAC using the infrastructure of the SLC positron source at the FFTB and FACET facilities. SLAC is currently planning for FACET-II to offer ability to test concepts in collider relevant regimes.

Plasma Wakefield Experiments at the SLAC final focus test beam (FFTB) facility (1998-2006) made the first measurements of acceleration and focusing in e$^+$ PWFA. Acceleration was observed at low beam/plasma densities with linear wakes~\cite{Hogan_1}. The beam evolution in meter long plasmas generates non-linearities and a large, non-Gaussian, beam halo is observed implying a large emittance~\cite{Hogan_2,Hogan_3}. Simulations show that the emittance grows rapidly along all longitudinal slices of the beam and make it clear that the e$^+$ beam evolution in the plasma is important for understanding the final beam parameters.

The FFTB was decommissioned to make way for the LCLS X-ray laser and the SLAC plasma wakefield programs relocated to FACET: a National User Facility that operated from 2012-2016. FACET preserved the SLC e$^+$ infrastructure and added a new compressor chicane that enabled compressed e$^+$ bunches for higher-densities and higher gradients: Q = 3.2\,nC/bunch, 1-10\,Hz, 20\,GeV, $\sigma_z \approx 20\,\mathrm{\mu m}$ (for e$^+$), $n_p \approx 10^{15}$ – $10^{17}\,\mathrm{cm}^{-3}$. Collimation techniques that were developed to make two tightly spaced e$^-$ bunches were used equally well for e$^+$ ones. This led to the first demonstration of controlled beam loading in the positron beam-driven wake and this scenario was tested in both the quasi-linear and non-linear regimes~\cite{Hogan_4}.

The hollow channel plasma is a structure that symmetrizes the response of the plasma to e$^-$ and e$^+$ beams. There is no plasma on axis, and therefore no focusing and, importantly for e$^+$, defocusing force from plasma ions. A hollow-channel plasma was created using a hollow laser beam in low density plasma with longitudinal wakefields that agreed with predictions~\cite{Hogan_5}. This concept works well when the system is perfectly symmetric but transverse misalignments and/or asymmetries in the beams give rise to strong transverse wakefields that were measured to be 10,000 times stronger than for CLIC~\cite{Hogan_6}.

FACET was able to provide high-density, compressed positron beams for non-linear PWFA experiments for the first time that yielded surprising new observations: within the plasma a single high-intensity e$^+$ bunch can bifurcate into two distinct energy populations with accelerated e$^+$ forming a spectrally distinct peak with an energy gain of 5\,GeV and an energy spread as low as 1.8\% (r.m.s.)~\cite{Hogan_7}. Numerical investigations provided insight into the mechanism. Some of the plasma e$^-$ can be pulled towards the axis and co-propagate with the e$^+$ beam loading both the transverse and longitudinal wakes.

Although a concept was developed for creating an electron-driver e$^+$-witness configuration at FACET, this was not realized before FACET was decommissioned to make way for the LCLS-II. The AAC community needs new facilities with beams suitable to test new concepts for e$^+$ acceleration in plasma. A plan was developed to restore e$^+$ capabilities to the FACET-II facility at SLAC with simultaneous delivery of up to 1\,nC e$^+$ and 2\,nC e$^-$ to the experimental area while also enabling the study of e$^+$ PWFA in electron wakes~\cite{Hogan_8}. Positrons were descoped from the FACET-II Project between 2016-2018 but User interest in e$^+$ did not fade away. The 2023 P5 report recognizes the importance of test facilities and in Area Recommendation 8 “…An upgrade for FACET-II e$^+$ is uniquely positioned to enable study of e$^+$ acceleration in high-gradient plasmas…”~\cite{Hogan_9}.

There are many new ideas in the literature that create and utilize the e$^-$ filament (first observed in experiments at FACET) to get acceleration and focusing of e$^+$ in plasma. Transversely tailored plasmas change the shape of the ionized plasma region modifying the trajectories of plasma e$^-$ in the wake. This leads to an elongated region in the back of the wake where e$^+$ bunches are focused and accelerated~\cite{Hogan_10,Hogan_11,Hogan_12,Hogan_13}. Transversely tailored drivers a.k.a. Wake Inversion creates a similar structure in the wakefield from a donut shaped e$^-$ drive beam~\cite{Hogan_14,Hogan_15}. Non-neutral fireball beams use the superposition of a co-propagating electron and positron beam to create a donut shaped charge distribution~\cite{Hogan_16}. High efficiency uniform wakefield acceleration of a e$^+$ beam using a stable asymmetric mode in a hollow channel plasma has shown promising results in simulation~\cite{Hogan_17}. If any of these techniques are to be deployed for collider applications, power efficiency will be critical and achieving a large accelerating gradient (even with good emittance) is not sufficient. A good comparison and survey of e$^+$ PWFA techniques may be found in G.J. Cao et al.~\cite{Hogan_18}.

Positron acceleration is 50\% of a plasma linear collider but only a small fraction of PWFA research. The non-linear blowout regime is great for e$^-$ but does not work for e$^+$. High-gradient acceleration of e$^+$ in plasma has been demonstrated, but alternative approaches engineering the plasma and/or beams to get all of the properties we want (gradient, efficiency, emittance…) need to be developed and tested. Research progress correlates with having the ability to test concepts experimentally. A plan has been developed to restore (and improve) our capabilities to test concepts for e$^+$ PWFA at FACET-II.

With the proposed e$^+$ upgrade FACET-II will be first facility capable of studying e$^-$-driven, e$^+$ witness PWFA. FACET-II will re-examine options with DOE HEP when response to P5 report is available. With a commitment and strong support from SLAC the plan could be executed on five-year time scale without interruption of existing user program.

Work supported in part by the U.S. Department of Energy under contract number DE-AC02-76SF00515.
\section{Interaction point physics in linear colliders based on laser-plasma accelerators \\ 
\small{\textit{T. Grismayer\textsuperscript{1}, W. Zhang\textsuperscript{1}, W. B. Mori\textsuperscript{2}, and L. O. Silva\textsuperscript{1}}}\normalsize}

\textit{\textsuperscript{1}GoLP/Instituto de Plasmas e Fus\~{a}o Nuclear, Instituto Superior T\'{e}cnico, Universidade de Lisboa, 1049-001 Lisboa, Portugal\\
\textsuperscript{2}University of California, Los Angeles CA, USA}

\subsection*{ Interaction Point Physics}

Interaction Point (IP) dynamics in linear colliders are critically important for high energy physics experiments. The dynamics at the IP e.g. due to collective processes can be deleterious to the beams, and can produce secondary particles that might hinder the outcome of the experiments. The most relevant collective beam phenomenon is disruption i.e. the effects of the electromagnetic fields of one beam on the opposing beam during extremely close encounters. This interaction can significantly affect the beam quality and the collision outcomes, making it a central for optimising the performance of linear colliders, such as the International Linear Collider (ILC) or the Compact LInear Collider (CLIC) \cite{Gris_Schulte2017}. 

\subsection*{Key Challenges}

\subsubsection*{Beam-Beam Interaction}
As the beams in a linear collider pass through each other at the interaction point, each beam experiences strong electromagnetic forces exerted by the opposite beam: $F_{\mathrm{beam}} \simeq m\omega_b^2\sigma_{\perp}$, where $\omega_b$ is the plasma beam frequency and $\sigma_{\perp}$ the bunch width. The equivalent laser strength parameter of the beam field is usually large compare to unity indicating an analogy with relativistic electron beam - intense laser scattering \cite{Gris_Fabrizio2019}. This interaction can cause the particles in the beam to deviate from their intended paths, leading to a phenomenon known as beamstrahlung where particles emit radiation due to transverse (with respect to the beam propagation direction) acceleration in the electromagnetic field of the opposing beam. This can result in a loss of beam energy and an increase in beam size. 

\subsubsection*{Disruption Parameter}
The disruption parameter is a crucial metric in linear collider design, quantifying the degree to which a beam is disrupted after passing through the opposing beam \cite{Gris_Chen1988}. It takes into account factors like the density of the beam particles, their energy, and the geometry of the beam collision. For future plasma-based collider, the typical beam size and energy will place the disruption parameter above unity \cite{Gris_Schroeder2023}. High disruption values often correlate with increased beamstrahlung and greater challenges in maintaining beam focus and stability \cite{Gris_Zhang2023}.

\subsubsection*{Beam Focusing and Stability}
Focusing the beams to extremely small sizes at the IP is essential to achieve high luminosity, which is indicative of the number of collisions per unit area per unit time. Advanced focusing techniques are employed to manage beam disruption and maintain the necessary beam stability and alignment. Plasma lenses have emerged as a novel and potentially revolutionary approach to focusing charged particle beams, offering several advantages over traditional magnetic lenses. These novel devices leverage the unique properties of plasma to influence and control particle beams.

\subsubsection*{Luminosity and strong field QED - disruption interplay}
Disruption impacts not only the quality of the beams but also the overall efficiency of collisions. Luminosity degradation due to higher levels of beamstrahlung and increased beam size can significantly affect the number of successful collisions and, consequently, the statistical significance of experimental results.
High-luminosity collisions between dense and high-energy (100s GeV to a few TeV) beams planned for future colliders,  will place beamstrahlung in the quantum regime where the QED parameter exceeds largely the unity, for which pair production cannot be neglected \cite{Gris_Zhang2023}. It was recently observed that disruption in the presence of strong field QED effects turns out to be a dynamical parameter that can be increased during the interaction. As a consequence, strong field QED should be carefully monitored to allow for the best luminosity as possible. 

\subsection*{Theoretical Implications and Future Prospects}
The theoretical understanding of beam disruption helps in designing better beam dynamics simulations and in predicting the outcomes of collisions in linear colliders. Practically, managing disruption involves a combination of advanced particle beam physics, strong field QED and plasma physics to optimize both the beam quality and the interaction point environment. As researchers continue to push the boundaries of what plasma-based linear colliders can achieve \cite{Gris_Schroeder2023}, understanding and mitigating IP physics with beam disruption becomes even more crucial. The physics of interaction points with disruption in plasma-based linear colliders encompasses a complex set of challenges and opportunities. Managing these disruptions is key to unlocking the potential of these colliders for high-precision experiments that probe the fundamentals of particle physics and the universe's deepest laws.
\vspace{1cm}
\section{Positron acceleration in plasma wakefields\\ 
\small{\textit{S. Corde}}\normalsize}

\textit{LOA, ENSTA Paris, CNRS, Ecole Polytechnique, Institut Polytechnique de Paris, 91762, Palaiseau, France}
\\
\\
The acceleration of positrons in a plasma is particularly challenging, especially so when targeting applications to high-energy physics (HEP) such as an electron-positron linear collider. Recently, we have published a review article in PRAB reviewing the field of positron acceleration in plasma~\cite{Cao24}, with a particular focus on the requirements of a collider. Although experimental facilities have been limited for the study of positron acceleration in plasmas, impressive progress has been made in the last 25 years on the experimental side, and more recently, extensive research has been carried out on the theoretical and numerical side. 

Looking back to what has been achieved, one can argue to some extent that one of the two main positron problems or challenges has been solved or can be considered as solvable. This first positron problem arises from the asymmetry of the nonlinear response of a plasma to positively- and negatively-charged particle beams, and finding a regime with a significant wakefield volume can \textit{simultaneously accelerate and focus a positron beam} can be challenging, as the blowout regime of electron acceleration cannot be straightforwardly used for positrons. Leveraging transverse beam loading, plasma and beam shaping, a plethora of schemes have been discovered with suitable properties for positron acceleration in plasmas~\cite{Cao24}. The second positron problem, that is to achieve the former with \textit{high quality and energy efficiency} as required for a collider, is yet to be solved, and is discussed in details in our review~\cite{Cao24} with a literature comparison of the performance in terms of luminosity per beam power $\mathcal{L}_P$, a key metric to assess the relevance of a scheme for the collider application. One of the critical conclusion from this analysis is that \textit{electron motion} is at the root of this second positron problem. Analog to the ion motion in the blowout regime of electron acceleration, it arises from the possible disruption of the flow of plasma electrons through the dense positron bunch, and can be quantified by the plasma electron phase advance in the positron bunch. A number of schemes, including loaded quasilinear or moderately nonlinear regimes, appear to be limited to a phase advance smaller than $\pi/2$, otherwise the plasma electron disruption strongly degrades the positron bunch quality. Different strategies exist to improve the luminosity per power, such as improving overall efficiency by energy recovery, using lower plasma focusing strength for the positron bunch or high Lorentz factor for plasma electrons, but one of the most effective strategy is to be resilient to large electron phase advances. An example is the finite-radius plasma~\cite{Diederichs23}, for which the resilience comes from the longitudinal spreading of plasma electron axis-crossing positions. Here plasma electrons do not flow through the positron bunch but cross the positron bunch even for a very strong positron load, each positron slice being focused by a different population of plasma electrons. With an improvement of about two orders of magnitude in $\mathcal{L}_P$ with respect to most other positron schemes, and now being only one order of magnitude away from the performance of CLIC or of electron acceleration in the blowout regime, these findings are opening very promising prospects in plasma-based positron acceleration research.

Looking ahead, there are two takeaways from this analysis: (i) a long-term intensive research is needed to uncover a fully self-consistent scenario for a plasma-based positron acceleration stage with the parameters required for a collider, including experimental demonstrations supporting such scenario, and (ii) targeting HEP applications of plasma accelerators that do not rely on plasma-based positron acceleration is critically important, given the highly uncertain outcome of the positron research in (i). Ideally, both (i) and (ii) should be pursued in parallel, yet (ii) is critical to support a credible plan for the application of plasma accelerators to HEP with much higher feasibility, and to fill the gap between today's state-of-the-art and the very challenging plasma-based electron-positron collider application.
\vspace{1cm}
\section{Generation and acceleration of polarised electron bunches in plasma accelerators\\
\small\textit{{K. P\~oder}}\normalsize}

\textit{Deutsches Electronen Synchrotron DESY, Notkestr. 85, 22607 Hamburg, Germany}
\\
\\ 
Spin-polarisation is a key aspect of any linear collider. Polarisation asymmetry can be used as an observable and opposite polarisations can enhance luminosity for certain studies. Particular polarisation combinations can enhance rates and suppress backgrounds. Any advanced collider concept must therefore ensure that the design is capable of preserving electron beam polarisation.  

The inherent compactness of plasma accelerators could enable huge savings in future linear colliders. To this end, multiple techniques to generate collider-level emittance and energy spread have been put forward. However, the topic of beam polarisation has only recently received more attention and multiple concepts to generate spin-polarised electron beams from plasma accelerators have now been put forward. A technique leveraging non-adiabatic multi-photon ionization in a beam-driven plasma wake was proposed by~\cite{Nie}
; the initial plasma source is completely unpolarized in this case. The polarization is limited by the ionization pathway asymmetry, with a 30\% polarization bunch of 3pC and 10-nm-scale emittance shown in simulations. Other proposed techniques make use of a pre-polarised plasma source, where the spins of all electrons are initially aligned. Such plasma sources can be prepared through molecular dissociation of suitably aligned hydrogen halide molecules~\cite{Spiliotis}
. Multiple proposals leveraging such a pre-polarised plasma exist~\cite{Wu}
. The most promising for near-term experimental demonstration is that from Bohlen et al~\cite{Bohlen}
, proposing to use colliding pulse injection to generate high quality, highly polarised electron beams. They apply Bayesian optimisation to demonstrate electron bunches with 30\,pC charge, 90\% polarisation and micron-level emittance from a laser plasma accelerator. This concept is currently the most promising pathway to an experimental demonstration of polarised electron beams from laser-driven plasma accelerator. Funding is being pursued to realise such a first demonstration at DESY. 

Work by Vieira et al~\cite{Vieira} 
examined beam polarisation preservation in high energy plasma accelerators. The key effect leading to beam depolarisation in a plasma accelerator is the precession of individual electron spins in the azimuthal focussing fields inside the plasma bubble. As the strength of these fields scales with distance from the axis, using smaller emittances and corresponding matched spot sizes in the plasma reduces beam depolarisation. Depolarisation up to 10\% was calculated for beams with tens of um emittances. Only percent level polarisation reduction was seen for zero-emittance beams for acceleration up to 1\,TeV. These encouraging results must be expanded upon, though, as that first study did not take multiple recently discovered effects into account. Such work is now becoming possible as reduced-model particle-in-cell codes that can simulate acceleration of 100-nm scale emittance in hundred-metre-scale plasmas (e.g. HiPACE++) also include spin precession physics. Leveraging these new tools, first simulations of beam polarisation preservation in a HALHF-type plasma accelerator are underway with results to be reported in the near future. 
\vspace{1cm}
\section{A few (interesting) aspects about collider modelling\\ 
\small{\textit{J. Vieira\textsuperscript{1}, P. Muggli\textsuperscript{2}}}\normalsize}

\textit{\textsuperscript{1}GoLP/Instituto de Plasmas e Fus\~{a}o Nuclear, Instituto Superior T\'{e}cnico, Universidade de Lisboa, 1049-001 Lisboa, Portugal\\
\textsuperscript{2}Max Planck Institute for Physics, 80805 Munich, Germany
}

\subsection*{Collider PIC simulations}


To maximize the luminosity in a particle collider, the %
beam %
normalized emittance needs to be kept as low as possible until the interaction point. For example, electron bunches 
for CLIC have vertical and horizontal emittances 
$\epsilon_{n,V} = 0.9~\mathrm{\mu m}$ and $\epsilon_{n,H} = 20~\mathrm{nm}$, with a corresponding 
growth 
budget of $\delta \epsilon_{n,V} = 0.01~\mathrm{\mu m}$ and $\delta \epsilon_{n,H} = 1~\mathrm{nm}$. These 
design parameters are challenging to account for, even 
in numerical %
simulation models. Thus, 
accurately modeling emittance evolution 
would be a major step forward for any plasma-based collider design, which may require systematic sensitivity studies to evaluate the accuracy of 
the models at the $nm$ %
emittance level. In the following, we outline a few interesting modelling aspects that arise by considering emittances and emittance tolerance budgets as low as those foreseen for CLIC.

For a round electron bunch in the nonlinear blowout regime of plasma wakefields, it is possible to minimize emittance growth by matching the bunch transverse size to the focusing force of the ion column, i.e., (see e.g., \cite{bib:vieira_joshi}):

\begin{equation}
\label{eq:sigma}
\sigma_{\perp} = \frac{1}{\sqrt{k_p}} \left(\frac{2 \epsilon_n^2}{\gamma}\right)^{1/4}\simeq1.3\times10^7\left(\frac{\epsilon_n^2 [\mathrm{\mu m}]}{E[\mathrm{GeV}] n_0[\mathrm{cm}^{-3}]}\right)^{1/4},
\end{equation}
where $k_p = \omega_p/c$ is the plasma wavenumber, $\omega_p=\sqrt{e^2 n_0/m_e \epsilon_9}$ is the plasma frequency, $n_0$ is the background plasma density, $m_e$ and $e$ are the electron mass and charge, $c$ the speed of light, and $\epsilon_0 = 8.85\times10^{-12}\,\mathrm{F/m}$ is the vacuum permittivity.

\begin{figure}[ht]
\centering
\includegraphics[width=1.0\textwidth]{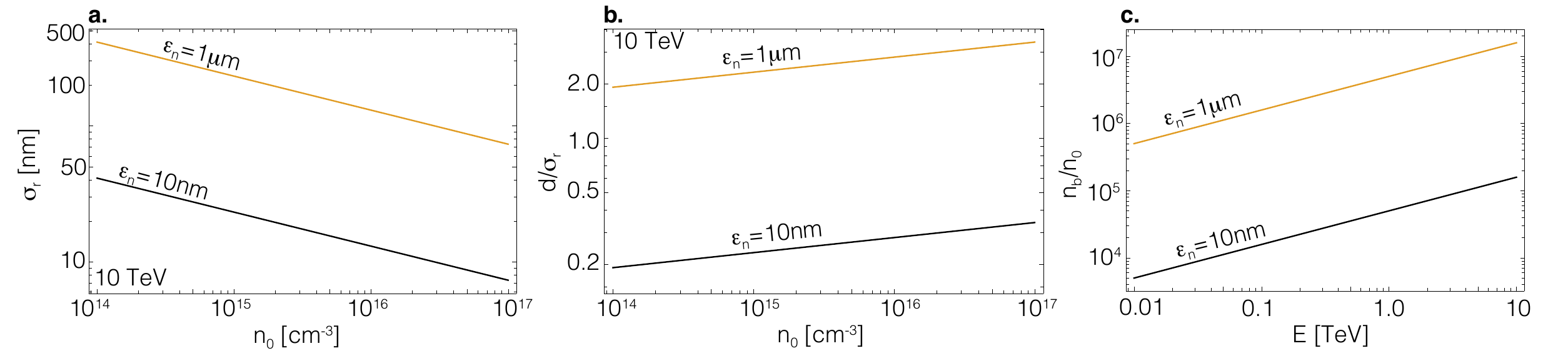}
\caption{\textbf{a.} Matched transverse size for a 10\,TeV electron bunch as a function of plasma density. \textbf{b.} Ratio between the transverse bunch size and inter-particle distance for a 10\,TeV electron bunch as a function of plasma density. \textbf{c.} Ratio between the bunch and plasma density as a function of bunch energy. %
All figures for two values of the normalized emittance of the bunch similar to those foreseen for CLIC.}
\label{figure1}
\end{figure}

Figure~\ref{figure1}a, shows the matched bunch 
size 
given by Eq.~(\ref{eq:sigma}) %
for two values of emittance. 
Properly resolving the %
very small transverse bunch size is an important factor for 
collider-relevant full PIC simulations, 
even with parameters similar to those of CLIC.

It is important to note that the ratio between 
$\sigma_{\perp}$
and the inter-particle 
distance ($d$) in the plasma, 
is approximately given by:
\begin{equation}
\label{eq:interpart}
\frac{\sigma_{\perp}}{d} \simeq 1.3 \frac{\epsilon_n^2 [\mathrm{\mu m}] n_0^{1/3} [\mathrm{cm}^{-3}]}{E[\mathrm{GeV}]}.
\end{equation}

Figure~\ref{figure1}b, 
shows from Eq.~(\ref{eq:interpart}) that in the low emittance plane of the beam, this ratio is always smaller than one. %
This suggests that understanding the influence of discrete particle effects in simulations %
(and in reality) may be important, particularly at lower plasma density. A similar consideration holds for the 
driver in laser wakefield accelerators (LWFA). In the LWFA, the inter-particle distance becomes comparable ($\lambda_0/d \lesssim 10$) to the 
wavelength %
of the pulse (assuming central wavelength $\lambda_0 = 1~\mathrm{\mathrm{\mu m}}$) for $n_0\lesssim 10^{15}~\mathrm{cm}^{-3}$.

Another important factor 
is ion motion, whose 
effect depends on the ratio between the electron bunch density and plasma density, given by:

\begin{equation}
\label{eq:ratio}
\frac{n_b}{n_0} = \frac{Q_b}{e V_b n_0}  \simeq 1.98\times10^5 \frac{Q\mathrm{[nC]} \sqrt{E\mathrm{[GeV]}}}{\alpha \epsilon_n[\mathrm{nm}]},
\end{equation}
where $V_b = \pi^{3/2} \sigma_{\perp}^2 \sigma_{\|}$ is the bunch volume (assuming a 
bi-Gaussian density distribution), $\sigma_{\|}=\alpha (2\pi/k_p)$ is the bunch length (i.e., $\sigma_{\|}$ is a fraction $\alpha$ of the plasma wavelength), and $Q_b$ its charge. 

Figure~\ref{figure1}c plots Eq.~(\ref{eq:ratio}) considering %
a witness bunch with $\epsilon_n = 1~\mathrm{\mu m}$ and $\epsilon_n = 1~\mathrm{nm}$. In 
this extreme regime, where $n_b/n_0 \simeq 10^4-10^6$, the influence of 
ion motion needs to be understood, especially during beam injection and extraction 
through plasma density ramps.

Finally, the tolerance budget may also 
apply to the uniformity of 
the wavefronts of the laser pulse, or %
of the 
transverse density distribution of the beam driver. As a rule of thumb, the tolerance on the laser-driver wavefront uniformity may be simply set as $\delta \epsilon_n \simeq L_{\mathrm{accel}}\delta k_{\perp}/k_0$, where $\delta k_{\perp}$ is a typical transverse wave-vector deviation from 
a flat front, $k_0=2 \pi/\lambda_0$ is the %
pulse central 
wavenumber, and $L_{\mathrm{accel}}$ is the laser propagation distance
 in plasma. Accordingly, for $L_{\mathrm{accel}}\simeq 1~\mathrm{m}$, and $\delta \epsilon_n = 1-1000~\mathrm{nm}$, $\delta k_{\perp}/k_0 \simeq 10^{-3}-10^{-6}~\mathrm{mrad}$.

\newpage





\chapterimage{images_head_icep}


\chapterimage{images_head_feuillep}
\glsaddall

\printglossaries

\addcontentsline{toc}{chapter}{Glossary}
\vfill\eject


\chapterimage{images_head_gravurep} 

\chapter{Committees and Participants}

 \textbf{Organizing committee} composed of Members of the ICFA ANA panel and of Jorge Vieira, Instituto Superior Técnico.

\vspace{0.5cm}
\noindent \textbf{List of workshop attendees}:
53 participants\\
Alexander Pukhov, University of Dusseldorf\\%
Alexei Kanareykin, Euclid Techlabs/Agronne National Laboratory\\
Angelo Biagioni, INFN\\
Arie Irman, Helmholtz-Zentrum Dresden-Rossendorf, Germany\\
Axel Huebl, Lawrence Berkeley National Laboratory\\
Bernardo Barbosa, IST - Lisbon\\
Bernardo Malaca, GoLP-IPFN of Instituto Superior Técnico\\
BoyuanLi,	Shanghai Jiao Tong University\\
Brian Foster,  University of Oxford/DESY\\
Brigitte Cros, LPGP-CNRS-Université Paris Saclay\\
Cameron	Geddes, Lawrence Berkeley National Laboratory\\
Camilla	Willim,GoLP / Instituto de Plasmas e Fus\~{a}o Nuclear Instituto Superior T\'{e}cnico, Lisbon, Portugal\\
Carlo Benedetti, Lawrence Berkeley National Laboratory\\
Chiara Badiali, GoLP/Instituto de Plasmas e Fus\~ao Nuclear, Instituto Superior T\'ecnico, Universidade de Lisboa, Lisbon, Portugal\\
Claudia	Romão, GoLP-IPFN of Instituto Superior Técnico\\
Denise Völker,	DESY\\
Edda Gschwendtner, CERN\\
Evan Ericson, PSI/EPFL\\
Francesco Massimo, LPGP - CNRS\\
Gianluca Sarri,	Queen's University Belfast\\
Ishay Pomerantz, Tel Aviv University\\
John Power, Argonne National Laboratory\\
John Patrick Farmer, MPP / CERN\\
Jorge Vieira, Instituto Superior Técnico\\
Jyotirup Sarma,Queens University Belfast\\
Kevin Cassou, CNRS/IN2P3/IJClab\\
Kristjan Poder, DESY\\
Laura Corner, Cockcroft Institute, University of Liverpool\\
Lucas Ivan, Inigo Gamiz	Instituto Superior Tecnico\\
Mark Hogan,	SLAC\\
Marlene	Turner, CERN\\
Massimo	Ferrario, INFN LNF\\
Michael	Backhouse, Imperial College London \\
Michele	Gallinaro, LIP Lisbon\\
Nelson Manuel Carreira Lopes, Universidade de Lisboa\\
Oznur Apsimon, The University of Manchester and the Cockcroft Institute\\
Patric	Muggli, Max Planck Institute for Physics\\
Philippe Piot, Northern Illinois University\\
Rafael	Russo de Almeida, GoLP, IST\\
Rajeev	Pattathil, Rutherford Appleton Laboratory\\
Riccardo Pompili, INFN\\
Richard	D'Arcy, University of Oxford\\
Roberto	Losito, CERN \\
Róbert	Babjak, GoLP - Group for Lasers and Plasmas,\\
Sebastien Corde, Ecole Polytechnique, LOA,\\
Severin	Diederichs,	CERN\\
Simon	Hooker, University of Oxford\\
Susanne	Schoebel, Helmholtz-Zentrum Dresden-Rossendorf\\
Thomas Grismayer, GoLP/IPFN, Instituto Superior Técnico, Universidade de Lisboa\\
Vera Cilento, CERN\\
Weiming	An, Beijing Normal University\\
Wim	Leemans, DESY\\
Zahra Mohammadzadehchamazkoti, GoLP/Instituto de Plasmas e Fus\~ao Nuclear, Instituto Superior T\'ecnico, Universidade de Lisboa, Lisbon, Portugal\\
\cleardoublepage

\end{document}